\documentclass[sigconf, table, prologue, screen]{acmart}

\usepackage[utf8]{inputenc}

\copyrightyear{2022} 
\acmYear{2022} 
\setcopyright{rightsretained} 
\acmConference[CHI '22]{CHI Conference on Human Factors in Computing Systems}{April 29-May 5, 2022}{New Orleans, LA, USA}
\acmBooktitle{CHI Conference on Human Factors in Computing Systems (CHI '22), April 29-May 5, 2022, New Orleans, LA, USA}
\acmDOI{10.1145/3491102.3517582}
\acmISBN{978-1-4503-9157-3/22/04}

\usepackage{acmart-taps}

\usepackage{xcolor}
\PassOptionsToPackage{table, prologue}{xcolor}
\usepackage{microtype}
\usepackage{color,soul}
\usepackage{xspace}
\usepackage{subcaption}
\usepackage{multirow}
\usepackage{verbatim}
\usepackage{enumitem}
\usepackage{tikz}
\usepackage{soul}
\usepackage{mathtools}
\newcommand{\note}[2]{{\color{#1}{[#2]}}\xspace}
\newcommand{\anon}[2]{#1}
\renewcommand{\anon}[2]{#2}

\newcommand{\colordot}[2]{{\textcolor{#1}{$\bullet$}}\emph{\underline{\bi#2}}\xspace}

\definecolor{clightgrey}{HTML}{80868b}
\definecolor{cgenerate}{HTML}{3B4D73}
\definecolor{cprompt}{HTML}{3c4043}

\definecolor{cexample}{rgb}{0.23, 0.30, 0.45}
\newcommand{\exinline}[1]{{\color{cprompt}``#1''\xspace}}
\newcommand{\quoteinline}[1]{{\color{cprompt}\emph{``#1''}\xspace}}

\newcommand{\eg}{\emph{e.g.,}\xspace}%
\newcommand{\ie}{\emph{i.e.,}\xspace}
\newcommand{\etal}{\emph{et al.}\xspace}

\newcommand{\modelname}{\tofix{\texttt{LaMDA}}\xspace}
\renewcommand{\modelname}{LLM\xspace}

\newcommand{\tprompt}[1]{{\color{cprompt}\texttt{#1}}\xspace}
\newcommand{\tgenerate}[1]{{\color{blue}\texttt{#1}}\xspace}
\newcommand{\tunit}[2]{\small\tprompt{#1}\tgenerate{#2}}

\newcommand{\taskquote}[1]{\newline
{
	\begin{minipage}{0.95\textwidth}
	\vspace{2mm}
	\noindent {\selectfont \color{darkgray}#1 \par}
	\vspace{2mm}
	\end{minipage}
	}
}
\newcommand{\squote}[1]{
    \hfill{
	\begin{minipage}{0.91\linewidth}
	\vspace{2mm}
	{\selectfont \emph{#1} \par}
	\vspace{2mm}
	\end{minipage}
	}
}
\newcommand{\paragraphBold}[1]{\paragraph{\emph{\textbf{#1}}}}
\newcommand{\sandbox}{\texttt{Sandbox}\xspace}
\newcommand{\chain}{\texttt{Chaining}\xspace}

\newcommand{\taskFlashcard}{\emph{Flashcard}\xspace}
\newcommand{\taskReview}{\emph{Review}\xspace}

\newcommand{\todo}[1]{\note{red}{\textbf{REVIEW}: #1}}
\renewcommand{\todo}[1]{}

\newcommand{\added}[1]{{\color{blue} #1}\xspace}
\newcommand{\removed}[1]{{\color{red} \sout{#1}}\xspace}
\renewcommand{\removed}[1]{}

\newcommand{\highlight}[1]{{\color{orange} #1}\xspace}

\renewcommand{\highlight}[1]{#1\xspace}
\renewcommand{\added}[1]{#1\xspace}

\newcommand{\revised}[1]{{\color{blue} #1}\xspace}
\renewcommand{\revised}[1]{{#1}\xspace}

\definecolor{cinput1}{HTML}{3182bd}
\definecolor{cinput2}{HTML}{6baed6}
\definecolor{coutput1}{HTML}{137333}
\definecolor{coutput2}{HTML}{1e8e3e}
\definecolor{cgenback}{HTML}{F1F3F4}
\definecolor{cgenfront}{HTML}{0076BA}

\newcommand{\conditionNameGrp}{[ \sandbox/\chain]\xspace}
\definecolor{coriginal}{HTML}{00BBD4}
\definecolor{cproblem}{HTML}{FF9701}
\definecolor{csuggestion}{HTML}{FFD932}
\definecolor{crewrite}{HTML}{4DAE50}
\definecolor{cgentext}{HTML}{0076BA}

\definecolor{cplace1}{HTML}{3182BD}
\definecolor{cplace2}{HTML}{137333}
\definecolor{ccity}{HTML}{00BBD4}
\definecolor{cinteraction}{HTML}{FF9701}
\definecolor{cinterest}{HTML}{FFD932}

\definecolor{cen}{HTML}{16E7CF}
\definecolor{cfr}{HTML}{4DAE50}

\definecolor{cvegalite0}{HTML}{00BBD4}
\definecolor{cdescribe}{HTML}{F27200}
\definecolor{callrule}{HTML}{0076BA}
\definecolor{crelaterule}{HTML}{FFD932}
\definecolor{cvalidresult}{HTML}{FF968D}
\definecolor{cvegalite1}{HTML}{4DAE50}

\definecolor{cconcept}{HTML}{00BBD4}
\definecolor{ctrait}{HTML}{FF9701}
\definecolor{cmetaphor}{HTML}{4DAE50}

\definecolor{cacronym}{HTML}{0076BA}
\definecolor{chasacronym}{HTML}{F27200}
\definecolor{cexpandcent}{HTML}{FFD932}
\definecolor{cfullsent}{HTML}{00AB8E}

\begin{document}

%%
%% The "title" command has an optional parameter,
%% allowing the author to define a "short title" to be used in page headers.
\title[Transparent and Controllable Human-AI Interaction by Chaining LLM Prompts]{AI Chains: Transparent and Controllable Human-AI Interaction by Chaining Large Language Model Prompts}

%%
%% The "author" command and its associated commands are used to define
%% the authors and their affiliations.
%% Of note is the shared affiliation of the first two authors, and the
%% "authornote" and "authornotemark" commands
%% used to denote shared contribution to the research.
\author{Tongshuang Wu}
\authornote{The work was done when the author was an intern at Google Inc.}
\email{wtshuang@cs.washington.edu}
%\orcid{0000-0003-1630-0588}
\affiliation{%
  \institution{University of Washington}
  %\streetaddress{P.O. Box 1212}
  %\city{Dublin}
  %\state{Ohio}
  \country{USA}
  %\postcode{43017-6221}
}
\author{Michael Terry}
%\authornotemark[1]
\email{michaelterry@google.com}
\affiliation{%
  \institution{Google Research}
  %\streetaddress{P.O. Box 1212}
  %\city{Dublin}
  %\state{Ohio}
  \country{USA}
  %\postcode{43017-6221}
}
\author{Carrie J. Cai}
\email{cjcai@google.com}
\affiliation{%
  \institution{Google Research}
  %\streetaddress{P.O. Box 1212}
  %\city{Dublin}
  %\state{Ohio}
  \country{USA}
  %\postcode{43017-6221}
}
%%
%% By default, the full list of authors will be used in the page
%% headers. Often, this list is too long, and will overlap
%% other information printed in the page headers. This command allows
%% the author to define a more concise list
%% of authors' names for this purpose.
%\renewcommand{\shortauthors}{Trovato and Tobin, et al.}

%%%% abstract.tex starts here %%%%

%%
%% The abstract is a short summary of the work to be presented in the
%% article.
\begin{abstract}
% v3
Although large language models (LLMs) have demonstrated impressive potential on simple tasks, their breadth of scope, lack of transparency, and insufficient controllability can make them less effective when assisting humans on more complex tasks.
In response, we introduce the concept of \emph{Chaining} LLM steps together, where the output of one step becomes the input for the next, thus aggregating the gains per step.
We first define a set of \emph{LLM primitive operations} useful for Chain construction, then present an interactive system where users can modify these Chains, along with their intermediate results, in a modular way. 
In a 20-person user study, we found that Chaining not only improved the quality of task outcomes, but also significantly enhanced system transparency, controllability, and sense of collaboration. 
Additionally, we saw that users developed new ways of interacting with LLMs through Chains: they leveraged sub-tasks to calibrate model expectations, compared and contrasted alternative strategies by observing parallel downstream effects, and debugged unexpected model outputs by ``unit-testing'' sub-components of a Chain.
In two case studies, we further explore how LLM Chains may be used in future applications.

\end{abstract}

%%%% abstract.tex ends here %%%%

%%
%% The code below is generated by the tool at http://dl.acm.org/ccs.cfm.
%% Please copy and paste the code instead of the example below.
%\begin{comment}
\begin{CCSXML}
<ccs2012>
<concept>
<concept_id>10003120.10003121.10011748</concept_id>
<concept_desc>Human-centered computing~Empirical studies in HCI</concept_desc>
<concept_significance>500</concept_significance>
</concept>
<concept>
<concept_id>10003120.10003121.10003129</concept_id>
<concept_desc>Human-centered computing~Interactive systems and tools</concept_desc>
<concept_significance>500</concept_significance>
</concept>
<concept>
<concept_id>10010147.10010257</concept_id>
<concept_desc>Computing methodologies~Machine learning</concept_desc>
<concept_significance>100</concept_significance>
</concept>
</ccs2012>
\end{CCSXML}

\ccsdesc[500]{Human-centered computing~Empirical studies in HCI}
\ccsdesc[300]{Human-centered computing~Interactive systems and tools}
\ccsdesc[100]{Computing methodologies~Machine learning}

%%
%% Keywords. The author(s) should pick words that accurately describe
%% the work being presented. Separate the keywords with commas.
\keywords{Human-AI Interaction, Large Language Models, Natural Language Processing}
%\end{comment}
%% A "teaser" image appears between the author and affiliation
%% information and the body of the document, and typically spans the
%% page.

%%%% abstract.tex starts here %%%%

%%
%% The abstract is a short summary of the work to be presented in the
%% article.
\begin{abstract}
% v3
Although large language models (LLMs) have demonstrated impressive potential on simple tasks, their breadth of scope, lack of transparency, and insufficient controllability can make them less effective when assisting humans on more complex tasks.
In response, we introduce the concept of \emph{Chaining} LLM steps together, where the output of one step becomes the input for the next, thus aggregating the gains per step.
We first define a set of \emph{LLM primitive operations} useful for Chain construction, then present an interactive system where users can modify these Chains, along with their intermediate results, in a modular way. 
In a 20-person user study, we found that Chaining not only improved the quality of task outcomes, but also significantly enhanced system transparency, controllability, and sense of collaboration. 
Additionally, we saw that users developed new ways of interacting with LLMs through Chains: they leveraged sub-tasks to calibrate model expectations, compared and contrasted alternative strategies by observing parallel downstream effects, and debugged unexpected model outputs by ``unit-testing'' sub-components of a Chain.
In two case studies, we further explore how LLM Chains may be used in future applications.

\end{abstract}

%%%% abstract.tex ends here %%%%

%%
\maketitle

%%
%% consistent spelling of the heading.

%%%% intro.tex starts here %%%%

\section{Introduction}
\label{sec:intro}

Large language models (LLMs) have introduced new possibilities for human-AI collaboration~\cite{bommasani2021opportunities}.
Pretrained on billions of inputs from the Internet~\cite{hendrycks2020measuring}, \revised{generative} models like GPT-3 can now perform a wide variety of tasks~\cite{bommasani2021opportunities}, ranging from translation~\cite{brown2020language}, to question answering~\cite{mishra2021natural}, and even advanced story writing~\cite{swanson2021story}.
These successes are enabled by their ability to adapt to desired tasks purely using \textbf{prompts}, or natural language descriptions of the tasks~\cite{reynolds2021prompt}. 
For example, one could adapt an LLM to act as a translation engine, simply by providing a few examples of the desired inputs and outputs: \exinline{English: How are you? French: Comment allez-vous? English: Hello! French:}
%a user could adapt an LLM to act as an English-to-French translation engine, simply by adding the keywords ``English'' and ``French'' to a prompt: \exinline{English: Hello! French:}
Based on this prompt, the model is likely to follow the pattern to output the correct French translation: \exinline{Bonjour!}
%Impressively, the prompts that trigger models' corresponding capabilities are in natural language, enabling an intuitive ``text in, text out'' iterative environment: a model outputs follow-ups on users' input text, and users in turn edit and feed them back to the model as a new round of input, starting the cycle again.

\begin{figure*}[t]
\centering
\includegraphics[trim={0 13.5cm 39cm 0cm}, clip, width=1\linewidth]{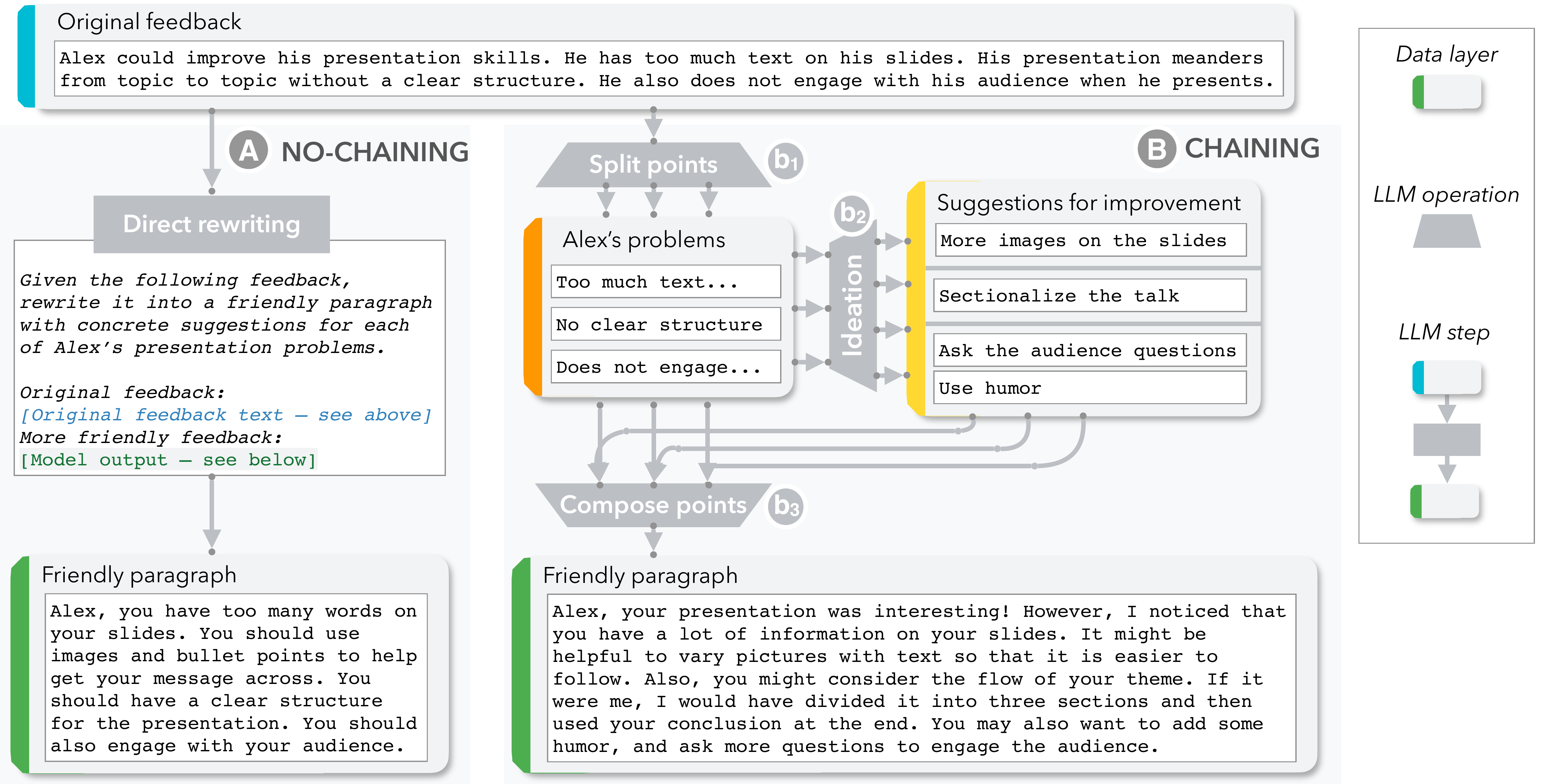}
\vspace{-15pt}
\caption{
A walkthrough example illustrating the differences between \emph{no-Chaining} (A) and \emph{Chaining} (B), using the example task of writing a peer review to be more constructive.
%A walkthrough example illustrating the differences between \emph{no-Chaining} (A) and \emph{Chaining} (B), using the example task of writing peer review paragraphs to be more constructive.
With a single call to the model in (A), even though the prompt (italicized) clearly describes the task, %\emph{describes the task} and \uline{identifies text sections}, 
the \colordot{crewrite}{generated paragraph} remains mostly impersonal and does not provide concrete suggestions for all 3 of Alex's presentation problems.
In (B), we instead use an LLM Chain with three steps, each for a distinct sub-task:
(b1) A \textbf{Split points} step that extracts each individual presentation \colordot{cproblem}{problem} from the \colordot{coriginal}{original feedback}, 
(b2) An \textbf{Ideation} step that brainstorms \colordot{csuggestion}{suggestions} per problem, and  %(split by the gray divider)
(b3) A \textbf{Compose points} step that synthesizes all the problems and suggestions into a final \colordot{crewrite}{friendly paragraph}. 
The result is noticeably improved.
}
\Description{A walkthrough example illustrating the differences between \emph{no-Chaining} and \emph{Chaining}, using the example task of writing a peer review to be more constructive.}
\vspace{-10pt}
\label{fig:flow_review}
\end{figure*}

The relative ease of natural-language-based prompt programming suggests that LLMs may be useful assistants for real-world tasks, with users customizing the models to their own needs.
In this light, recent work in Natural Language Processing (NLP) has begun to examine the algorithmic capabilities of LLMs, mostly on synthesized tasks~\cite{wang2021exploring, floridi2020gpt, reynolds2021prompt}.
However, many real-world tasks can be quite complex (\eg outlining long essays, debugging software code), and may present challenges for current LLMs to solve from a single model run.
For example, as LLMs learn the forms of language~\cite{bender2020climbing}, they produce lower quality outputs when solving tasks that require multi-step reasoning~\cite{tan2020progressive, branwen2020gpt, wei2022chain}.
%due to their limited ability to find, refine, and connect \emph{multiple} function blocks~\cite{branwen2020gpt}, LLMs produce lower quality outputs when solving tasks that require multi-step reasoning~\cite{tan2020progressive, ranzato2015sequence}.
Likewise, they may fail to capture the subtleties of many tasks that involve multiple objectives simultaneously \revised{(\eg identifying \emph{and} fixing multiple bugs in a code snippet)}.
\todo{Introduction: Provide examples of tasks that involve multiple objectives simultaneously.}
%Figure~\ref{fig:flow_review} shows a task in which the goal is to rewrite peer feedback to be more friendly, with concrete suggestions.
\added{Figure \ref{fig:flow_review} shows a task involving multiple concurrent objectives: (1) to rewrite peer feedback to be more friendly, and (2) to rewrite it with additional concrete suggestions, and (3) to ensure that each noted sub-problem (\eg too many words on slides, presentation meaders, does not engage with audience) is addressed.}
% As can be seen in Figure~\ref{fig:flow_review}A,
While an LLM can both generate suggestions~\cite{gpt-idea-machine} and adjust the tone in isolation (\eg in \cite{stylerecipe}), it lacks the capability to perform both tasks together well in an end-to-end manner. 
As a result, it produces a mediocre paragraph that only meets a few requirements (see output of Figure~\ref{fig:flow_review}A).

%Moreover, the powerful, open-ended scope of LLMs can simultaneously make them opaque and difficult to debug. 
%Because LLMs can essentially take any natural language as input, end users may find it difficult to decipher which specific edits need to be made to remedy unexpected model outputs or to establish an accurate mental model of LLM capabilities and limitations. 
%There are no obvious edits that can, for instance, encourage the model to provide more suggestions related to ``too much text on slides'' in Figure~\ref{fig:flow_review}A.

Besides being inherently limited for complex problems, LLMs are also difficult to interact and collaborate with, as they can be opaque and hard to debug.
Since LLMs can take in any natural language prompts, end users may struggle to determine how to change their prompts to remedy unexpected model outputs. 
They may also have difficulties developing accurate mental models of an LLM's capabilities and limitations.
%It may also be difficult for them to develop an accurate mental model of an LLM's capabilities and limitations.
There are no obvious edits on the prompt that can, for instance, encourage the model to add more suggestions regarding \exinline{too much text on slides} in Figure~\ref{fig:flow_review}A.
%Unfortunately, less work has explored how LLMs can better support human activities interactively.
%\sherry{here}

In this work, we introduce the notion of \textit{Chaining} multiple LLM prompts together, to help users accomplish complex tasks with LLMs in a way that is more transparent and debuggable.
Chaining takes advantage of LLMs' unique ability to handle a variety of independent tasks.
In a Chain, a problem is broken down into a number of smaller sub-tasks, each mapped to a distinct step with a corresponding natural language prompt; results of one or more previous steps are aggregated in the next step's input prompt.
%Results from one or more steps are aggregated into the input prompt for the next step. 
Thus, Chaining enables users to run \emph{the same} model on multiple sub-tasks, thereby granting each sub-task a higher likelihood of success (as opposed to solving the entire task in one go).
%Thus, Chaining enables users to run \emph{the same} model on multiple sub-tasks, where each sub-task has a higher likelihood of being solved well by the model 
%(relative to the model performance if it attempted to solve the entire problem with a single prompt and model run).
%Thus, rather than applying a single run of an LLM on a complex task that it may not perform well on, Chaining enables people to run the \emph{same} model on multiple sub-tasks that it is more likely to perform well on, and accumulate those gains across multiple runs of the model.
%In Figure~\ref{fig:flow_review}B, by first extracting unique presentation problems ($b_1$) and ideating suggestions per problem ($b_2$), the final composed ($b_3$) paragraph is noteably more friendly and constructive, even though the underlying LLM remains the same.
In Figure~\ref{fig:flow_review}B, while the underlying LLM remains the same, by \emph{splitting} (\ie, extracting) presentation problems ($b_1$) and \emph{ideating} suggestions per problem ($b_2$), the final \emph{composed} paragraph ($b_3$) is more comprehensive in addressing all problems, and has a more constructive tone.

In addition to potentially improving outcomes, Chaining opens up new channels for fine-grained human feedback and control. 
For example, thanks to the separate \emph{Ideation} step in Figure~\ref{fig:flow_review}$b_2$, Chaining allows users to customize which suggestions to include in the final paragraph, an operation that is unavailable without Chaining (Figure~\ref{fig:flow_review}A).
%We provide an interactive interface to help users selectively improve parts of the solution. 
We develop an interactive interface to expose these additional ``knobs'' to end users.
The interface visualizes the Chain structure, and allows users to customize a Chain at various levels:
they can iterate on the local prompts in each step, edit intermediate data between steps, or modify the entire Chain.
To inform the design of this tool, we surveyed 73 existing LLM use cases and summarized them into a set of \emph{LLM primitive operations}, each with default prompting and data structures. 
They help inform what types of sub-tasks could be used within a Chain, as well as how those steps can feed into each other.
%They help inspire effective task decomposition, and how steps can feed into one another.

To evaluate the impact of Chaining on both task performance and user experience, we conducted a within-subject user study, in which 20 participants completed tasks using both Chaining and a standard (non-Chaining) interface\added{, with the same underlying LLM powering all the steps in the Chaining interface, as well as the non-Chaining one.}
Our results show that Chaining significantly improved key dimensions of the human-AI experience: transparency, controllability, collaboration, and mental support. In addition, participants also achieved higher-quality outcomes $\sim$82\% of the time using Chaining.
We also saw participants leveraging Chaining for purposes beyond immediate task accomplishment --- they calibrated their expectations of the model using the smaller scope of sub-tasks,
explored alternative prompting strategies by comparing parallel downstream effects, 
and debugged unexpected model output by isolating and ``unit-testing'' different parts of a Chain.
\highlight{Critically, these improvements were achieved \textit{without changing the model itself}. }
These findings suggest that one way to improve the explainability and debuggability of an otherwise opaque, black-box LLM is to have it \textit{do less}: breaking a problem up into smaller problems, having the model solve each (smaller) problem separately, showing the intermediate results, and allowing users to edit those results.

\added{The ability to chain LLM calls using a set of Chaining building blocks, within an interactive interface, collectively represents a novel method and system for prototyping new AI-powered tasks and features using LLMs.}
We conclude the paper with case studies illustrating how Chaining can support more diverse applications in the future, as well as insights into challenges and opportunities that arose from our experiments.
%For example, we envision Chaining to generate intermediate data points that help justify its final output, and support rich prototyping on AI-infused applications.
% Finally, we illustrate a diverse set of applications that Chaining could be used for in the future. 
% Through case studies on visualization bug fixing and shorthand expansion, we envision that Chaining can help provide rationales in human-AI decision making tasks where the gold answer is unclear, and support rich prototyping on AI-infused applications.
% We conclude the paper with discussions on opportunities and additional challenges emerged in our studies, including using LLMs to help write LLM instructions to increase prompting diversity, quantifying the impact of cascading errors, etc.
In summary, we contribute:
\begin{itemize}[leftmargin=1.2em,labelwidth=*,align=left]
\item We introduce the notion of \emph{LLM Chaining}. Through a series of chained model calls, each targeting a small and well-scoped sub-task, we adapt \emph{a single LLM} to contribute to \emph{multiple sub-components} of a task.
\item We design and implement building blocks for constructing and interacting with LLM Chains. 
These include a set of primitive LLM operations representing functions well-scoped for a single model run, and an interactive interface that displays the intra- and inter-step structures of a Chain. 
Users can run Chains step-by-step, and customize them at various granularities (editing intermediate model outputs, rewiring steps, etc.). 
% Can you be more specific than "various granularities"?
\item We report results from a 20-person evaluation that shows Chaining can increase system transparency, controllability, and task outcomes. 
Importantly, these gains are achieved without any changes to the underlying model. 
Combined with the case studies, we demonstrate the potential of improving explainability and debuggability of LLMs through task decomposition and finer-grained application of LLM models. %exposing and structuring various interaction strategies. 
\end{itemize}

Taken together, our findings inform the design and research of future human-LLM collaborative systems, an area of critical importance in years to come.

\section{Background and Related Work}
\label{sec:relate}

\subsection{Large Language Models}
\label{ssec:relate-llm}

\revised{A generative language model is primarily designed to continue its input with plausible output (\eg given a prompt \exinline{I went to the}, it might auto-complete with \exinline{coffee shop}). 
However, when pre-trained on billions of samples from the Internet, recent transformer-based LLMs~\cite{vaswani2017attention} like GPT-3~\cite{brown2020language} and Jurassic-1~\cite{J1WhitePaper} encode enough information to support additional \emph{in-context learning}: they can be easily customized at run time (without any re-training needed) to handle new tasks beyond text continuation.
To invoke the desired functionality, users write natural language instructions, or \emph{prompts}~\cite{lu2021fantastically, betz2021thinking, liu2021makes}, that are appropriate for the task. 
The most common patterns for prompting are either zero-shot or few-shot prompts. 
Zero-shot prompts directly describe what ought to happen in a task. 
For example, we can enact English-to-French translation with a prompt such as \exinline{Translate the sentence ``Do you like the weather?'' to French:}.
In contrast, few-shot prompts show the LLM what pattern to follow by feeding it examples of desired inputs and outputs: \exinline{[English] Hello! [French] Bonjour! [English] Do you like the weather? [French]}.
Given either of these prompts, the LLM may respond with the French translation \exinline{Vous aimez le temps?}~\cite{jozefowicz2016exploring}.
Importantly, such task customization happens on the fly and, as a result, a single LLM can be flexibly adapted to a wide variety of use cases like code generation, question answering, creative writing, etc.~\cite{brown2020language, swanson2021story}.
This flexible adaptation, together with the text-in, text-out structure, creates an intuitive natural language interface between humans and the model.}

Despite their versatility, LLMs require careful prompt design.
Various studies therefore focus on \emph{prompt engineering}~\cite{lu2021fantastically, betz2021thinking, liu2021makes}.
As manual prompting can be sub-optimal, some work automatically mines more effective prompts. %by finding demonstrative cases (called few-shot examples) or prompt templates. 
However, the mined prompts tend to be less human-readable~\cite{shin2020autoprompt} and therefore less compatible with human-AI interaction.
Conversely, strategies like progressive generation (\ie multi-round text expansion)~\cite{tan2020progressive} and meta-prompting (\ie asking the model to elaborate on the problem)~\cite{betz2021thinking, reynolds2021prompt} attempt to seed LLMs to generate more effective prompts before solving the task.
%For example, the serialization meta-prompt \exinline{split this problem into steps} encourages the LLMs to automatically decompose a given task.
In essence, these approaches also adopt the spirit of multi-step problem solving, but focus on expanding the context \textit{without} human intervention.
%all target %at the subset of 
%\emph{accumulating} information 
%In contrast, our work more comprehensively defines the scope of task Chaining, with primitive operations that highlight various LLM functionalities, Chained steps that can not only add but also remove information along the LLM Chain, and fixable intermediate data points exposed.
Our work defines Chaining more comprehensively, with primitive operations that illustrate LLM capabilities, LLM steps that can add or remove information along the Chain, and editable intermediate data points.

%Instructions and prompts curated per task are usually more effective~\cite{reynolds2021prompt}, and we observe diverse prompt strategies for the same task throughout our surveys (\eg question answerings with chatbots, purely rely on descriptive prefixes, etc.)

%%%%%%%%%%%%%%%%%%%%%%%%%%%%%
\subsection{Human-AI Collaboration}

Human-AI interaction has been explored in domains such as classification~\cite{bansal2021does, smith2021no}, drawing~\cite{oh2018lead, davis2016empirically}, translation~\cite{green2014predictive}, creative writing~\cite{clark2018creative, gero2019metaphoria}, and design ideation~\cite{koch2019may}.
%The vision is that AI systems should augment or enrich human collaborators, rather than replace them, based on the current human inputs.
Prior work has noted core challenges of the interaction, such as a lack of transparency, controllability, and user agency~\cite{amershi2019guidelines, buschek2021nine,huang2020ai}.
Through Chaining, we aim to address these user-centered concerns.

In a collaboration, AI can play various roles, such as casual creators that encourage exploration~\cite{davis2016empirically} or assistants that compensate for human weaknesses~\cite{levy2021assessing, wu-etal-2021-polyjuice}.
For example, Gero \etal~\cite{gero2019metaphoria} showed that generators could serve as cognitive offloading tools so that humans could focus their attention where it is needed most, a core motivation that we also share. 
Cai \etal~\cite{cai2019human} investigated how a medical AI can assist with doctors' decision-making process during prostate cancer diagnosis, by helping them compare and contrast similar images. 
Most of these studies, however, use task-specific models, and therefore limit observations to human interaction with AI that primarily serves one function, or in one domain (\eg writing, medicine, music, etc.). 
DuetDraw~\cite{oh2018lead} may be an exception to this, as it uses several models, each of which supports a different co-drawing functionality. 
Rather than training multiple models for different tasks, or using a single model for a single type of task, our work explores how a \textit{single} large language model (with inherently customizable capabilities) can support humans in \emph{a variety of sub-tasks}. 
Finally, the closest work to ours might be online interfaces for users to interactively create prompts\footnote{\url{https://gpt3demo.com/apps/openai-gpt-3-playground}}, or interfaces enabling users to perform natural language programming of code using a large language model~\cite{jiang2021genline}.
These systems used prompt engineering to create a set of programming-related functionality for users. While this prior work focused on single prompts, our work looks at how Chaining \textit{multiple} prompts can address a much wider range of human tasks, and evaluate its effects on user experience.  
%DuetDraw~\cite{oh2018lead}, a co-drawing tool that involve multiple generators, each supporting a subset of five helper functions.
%However, DuetDraw users only adopt these models separately for finishing incomplete objects, colorization, and other tasks, whereas data-sharing and aggregation are central to our Chaining.
%Nevertheless, our design is inspired by their observation that people preferred to make their own decisions, and would respond passively when the models' contribution did not correspond with their task goal, or when they had limited capabilities curating model outputs.

%%%%%%%%%%%%%%%%%%%%%%%%%%%%%
\subsection{Workflows in Crowdsourcing}
\label{ssec:relate-crowdsourcing}

Though less prevalent in human-AI collaboration, the concept of Chaining is inspired by concepts of ``pipelining'' and ``microtasking,'' which have long been used in crowdsourcing~\cite{cai2016chain,teevan2016productivity}. 
In crowdsourcing, requesters break down complex tasks into pieces that can be performed independently, then combined~\cite{chilton2013cascade, kim2017mechanical, law2011towards, retelny2017no}.
Previous research shows that decomposed tasks allow the completion process to become more structured~\cite{chilton2016humortools} and more resilient to interruptions~\cite{cheng2015break}, something we also witness in our user study.
The goal of crowd workflows is typically to address and safeguard against the limitations of a typical worker.
For example, Bernstein \etal~\cite{bernstein2010soylent} ensured text editing quality through a Find-Fix-Verify workflow, which modulates the scope of sub-tasks to reduce variance of crowdworker effort.
%, and increase redundancy.
Meanwhile, Context Trees~\cite{verroios2014context} hierarchically summarize and trim the otherwise overwhelming global contexts, making them compact enough for a single worker to digest.

Our Chaining approach also aims to address pitfalls of a single LLM pass, but the pitfalls are somewhat distinct. 
While crowdsourcing focuses more on cognitive load and task duration --- factors that can affect the performance of human workers~\cite{kulkarni2011turkomatic} --- for LLMs with intensive computing power, their limitations err towards a lack of reasoning abilities, high variance of prompt effectiveness, and exposure bias.
A thorough analysis of these AI issues is needed for constructing and chaining LLM steps, which we illustrate in Section~\ref{ssec:llm-challenge}, and address through the design of primitive operations in Table~\ref{table:primitive_category}.
Through user studies (Section~\ref{sec:user-study}) and case studies (Section~\ref{sec:case-study}), we demonstrate that Chaining can effectively address these issues. Finally, our work also shares challenges found in crowdsourcing workflows, such as handling cascading errors that affect later stages~\cite{kittur2011crowdforge}, staged crash-and-rerun~\cite{little2010turkit}, all of which we take into consideration in the design of the Chaining structure. %and the interactive interface.
%For example, we ask humans to fix intermediate errors, rather than LLM ``workers,'' whereas the switch between serialization and parallelism is handled through the data structures across operations (in Section~\ref{ssec:primitive}). 
Beyond this, we advance the field by examining how core features of Chaining (\eg cascading effects, parallel paths) are used not only to accomplish tasks, but also to aid in increasing the transparency and debuggability of AI.
%the switch between parallel (\eg ideating suggestions per problem in Figure~\ref{fig:flow_review}) and iterative processes (rewriting sentences)~\cite{little2010exploring},

%%%% relate.tex ends here %%%%

%%%% definition.tex starts here %%%%

\section{Chaining LLMs}
\label{sec:pipeline}
%Although there have been numerous impressive demonstrations of LLMs doing highly targeted tasks well (\eg aiding with email responses~\cite{thiergart2021understanding}, detecting hate speech~\cite{chiu2021detecting}, or extracting plans~\cite{olmo2021gpt3}),
%there are many contexts in which LLM performance may suffer, such as if the data is formatted differently, if there is extraneous data in the input, if the task inherently demands multiple sub-parts, or if the user is asking the model to perform several tasks at once.
%Given this observation, we hypothesize that decomposing a problem into smaller, highly targeted tasks is likely to increase model performance on those sub-tasks, and by extension, the overarching task.

Despite the impressive capabilities of LLMs, there may be contexts in which LLM performance would suffer, such as if the data is formatted sub-optimally, if there is extraneous data in the input, if the task inherently demands solving multiple sub-parts, or if the user is asking the model to perform several tasks at once. 
Meanwhile, LLMs may perform \emph{highly targeted} tasks well.
By narrowing the scope and context of an LLM operation, for example, LLMs may themselves be useful for addressing some of their own challenges (\eg removing extraneous data, splitting problems into sub-parts, etc.). 
Thus, we hypothesize that decomposing a problem into smaller, highly targeted tasks is likely to increase model performance on those sub-tasks, and by extension, the overarching task. 
%As a result of accumulating the gains in performance found by solving smaller task, Chaining can solve tasks that would be difficult for a single run of an LLM in a single pass.

We define \textit{Chaining} as the process of breaking up complex tasks into smaller steps, where each step can be completed by an independent run of an LLM, and where the output of one or more steps is used as input for the next. To identify tasks that are most likely to benefit from Chaining, we first surveyed existing language modeling literature, and summarized \textbf{common challenges} LLMs face.
\added{As described in Section~\ref{ssec:llm-challenge}, these challenges are caused by the underlying modeling structure \emph{shared} by the mainstream LLMs, including but not limited to GPT-3, Jurassic-1, and the internal LLM used in Section~\ref{sec:user-study} and~\ref{sec:case-study}.}
Then, to identify promising sub-tasks that could be used as building blocks, we surveyed existing online demos of LLMs, and curated a list of \textbf{primitive LLM operations}, which may help overcome those challenges by scoping the inputs/outputs to be more amenable to what an LLM can handle.

%Thus, we define \textit{Chaining} as the process of breaking up complex tasks into smaller steps, where each step can be completed by an independent run of an LLM, and where the output of one or more steps is used as input for the next. 
%As a result of accumulating the gains in performance found by solving smaller task, Chaining can solve tasks that would be difficult for a single run of an LLM in a single pass.
%Chaining uses crowdsourcing as an inspiration (reviewed in Section~\ref{ssec:relate-crowdsourcing}), which seeks to decompose difficult tasks into smaller ones that are easier for human workers to accomplish.
%likewise, Chaining is best suited to tasks that are challenging for LLMs as a whole, but that can be achieved through a series of smaller tasks an LLM is adept at completing.

%To identify LLM strengths and weaknesses, we surveyed existing language modeling literature and online LLM demos.
%We summarize common challenges LLMs face, and curate a set of \textbf{primitive operations} to help address them, which represents the task space that appear to be well-suited for a single run of an LLM.

%%%% figures/primitives_two_column.tex starts here %%%%

\begin{table*}[t]%1
\centering
\begin{tabular}{@{}c@{}}
\includegraphics[width=\textwidth]{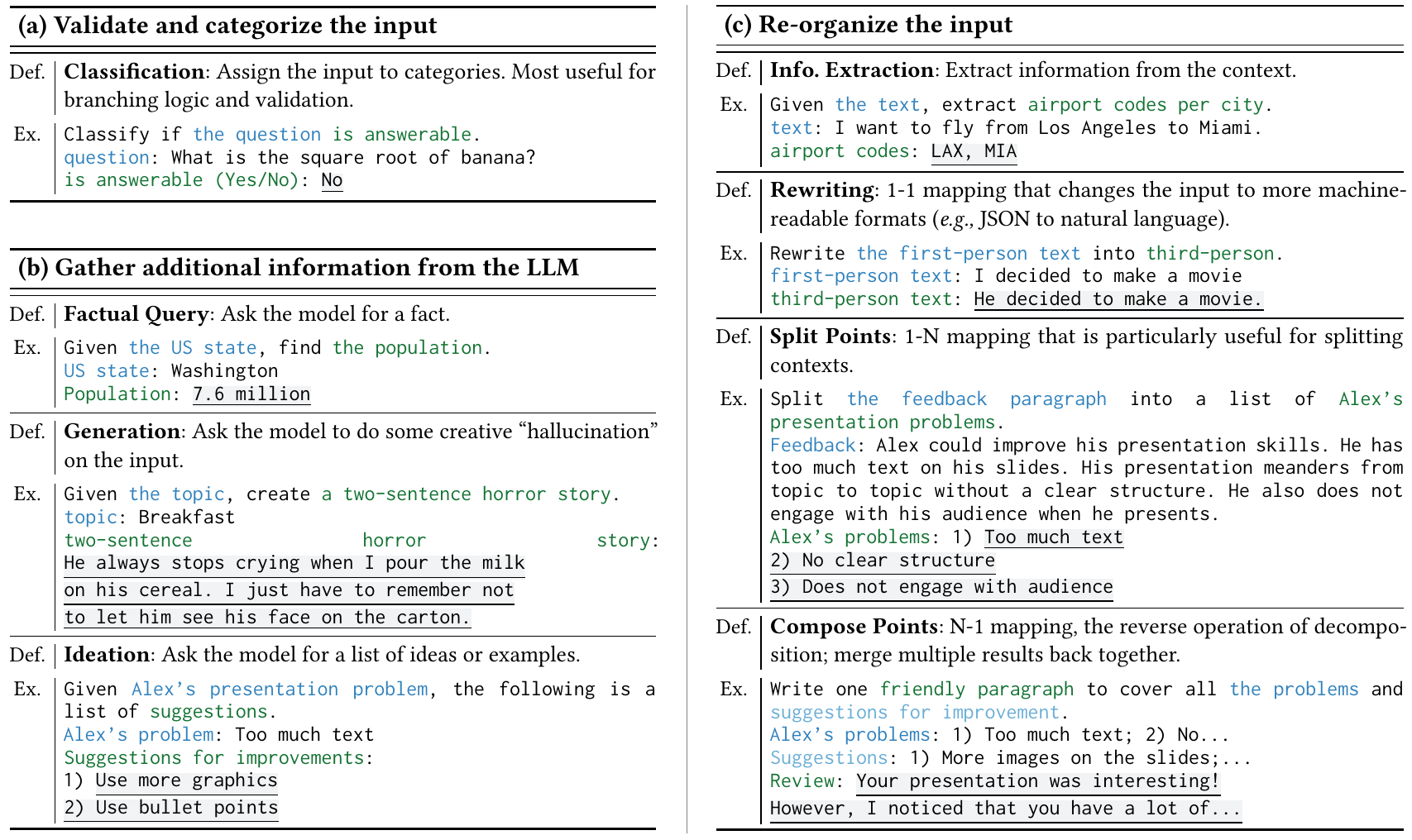}
\end{tabular}
\label{table:primitive-exam-context}
\vspace{\anon{0pt}{5pt}}
\caption{We curate eight primitive operations that may be adequately handled by a single LLM run.
Grouped according to their intended objectives, these operations can help address the LLM challenges detailed in Section~\ref{ssec:llm-challenge}.
Along with the definitions, we provide examples of prompts that enact these operations, with the \underline{underlined text} being the LLM output given the preceding prompt.
The examples for \emph{Ideation}, \emph{Split} and \emph{Compose} points are replicas of steps in Figure~\ref{fig:flow_review}.
The full implementations (with the parameters in Figure~\ref{fig:log_distribute}) are in Appendix~\ref{sec:appendix-primitive-implement}.}
\Description{The table describes three types of primitive operations that may be adequately handled by a single LLM run.
Along with the definitions, we provide examples of prompts that enact these operations, with the underlined text being the LLM output given the preceding prompt.}
\label{table:primitive}
\vspace{\anon{-20pt}{-15pt}}
\end{table*}

%%%% figures/primitives_two_column.tex ends here %%%%

%\input{figures/primitives}

\subsection{LLM Challenges \& Primitive Operations}
\label{ssec:llm-challenge}

Existing literature exposes three main challenges that LLMs face:
%, or scenarios LLMs are \emph{less capable of} handling:

% wang2021exploring: We guess that the model does not have calculation ability, but rather remembers some examples that have appeared before, since each calculation with 1000 and ‘12345678 + 87654321’ may appear in the Internet for many times while ‘12345678 + 8765432’ may not so frequently appear
% betz2021thinking: Pre-trained neural language models, whether uni- or bidirectional, display a poor zero-shot performance on reasoning tasks
\begin{enumerate}%[nosep,labelwidth=*,leftmargin=1.8em,align=left,label=C.\arabic*]
\item[C.1] %\label{challenge:reasoning} 
\textbf{LLMs lack multi-step reasoning capabilities}.
Because LLMs are designed to grasp the \emph{form} of language, rather than the \emph{meaning}~\cite{bender2020climbing}, they can struggle on tasks like sequential arithmetic problems, multi-hop question answering, recognizing and comparing sentences, or those that require branching logic~\cite{betz2021thinking, floridi2020gpt, wang2021exploring, branwen2020gpt, wei2022chain}.

% tan2020progressive: GPT-3 (Brown et al., 2020) was reported to produce long essays, yet the results seem to need extensive human curations (e.g., MarketMuse; Gardian), and the system is not publicly available to adapt to arbitrary desired domains.
\item[C.2] %\label{challenge:exposure-bias} 
\textbf{LLMs suffer from \emph{exposure bias}}~\cite{tan2020progressive, ranzato2015sequence}.
Because LLMs generate text sequentially in an autoregressive manner (the tokens generated by the models are themselves used to predict the next word), errors or imperfections from previous runs can accumulate. 
Thus, LLMs are less likely to perform well when generating long bodies of text.
%An workaround might be to use two steps in Table~\ref{table:primitive-format-context}: \ie \emph{split} the tasks at hand, perform each task with shorter output size (\eg rewriting at sentence levels) and re-\emph{compose} the results afterwards, simulating the progressive generation idea from Tan \etal~\cite{tan2020progressive}.
Exposure bias can also cause LLMs to produce redundant content, in some severe cases repeating the same phrase over and over again~\cite{holtzman2019curious, welleck2020consistency}.
As a result, they struggle to generate text with diverse themes or arguments (\eg suggestions for all three problems in the peer review example in Figure~\ref{fig:flow_review}).

\item[C.3] %\label{challenge:prompt} 
\textbf{LLMs are sensitive to input prompts.}
They tend to favor certain prompt formats, paraphrases~\cite{o2021context, lu2021fantastically}, or even certain information in the input.
%their artificial nature makes them unlikely to appearin the pre-training data
%For example, unnatural text tends to be less efficient~\cite{branwen2020gpt}, while nouns and verbs are more important than adjectives and function words~\cite{o2021context}.
For example, prompts that are unnatural relative to the typical text distribution tend to be less efficient~\cite{branwen2020gpt}, while nouns and verbs are more important than adjectives and function words~\cite{o2021context}.
\end{enumerate}

%Meanwhile, there are numerous online demos using LLMs that illustrate their unique strengths and capabilities, such as aiding with email responses~\cite{thiergart2021understanding}, detecting hate speech~\cite{chiu2021detecting}, or extracting plans~\cite{olmo2021gpt3}.
%These strengths could potentially be used to address LLMs’ own weaknesses.

These challenges tend to stem from tasks being too broad. Yet, as discussed above, LLMs may be able to perform certain tasks well if they are highly targeted, with narrower contexts. 
\todo{Page 6: Where do the 73 existing demos come from and how were they selected?}
Hence, with these challenges in mind, we reviewed 73 existing demos \added{based on an extensive search of official LLM websites, social media, and published case studies (these are enumerated in Table~\ref{table:primitive_category}, Appendix~\ref{sec:appendix-primitive-category})} to identify promising LLM capabilities that may help scope the inputs/outputs, culminating in a set of primitive operations.
Note that the operations we identified may not be exhaustive, but rather represent an interesting range for study, with a variety of operations addressing each LLM challenge. 
Pilot studies --- as well as use cases we present later --- suggested these were a reasonable set to pursue. 
\highlight{Full details of our methodology can be found in Appendix~\ref{sec:appendix-primitive-category}}. 
%We note that our set of operations is similar to those identified in the GPT-3 tutorial page, and that our use of these primitives in our research also suggests that these provide a reasonable set of core operations for Chaining.

Table ~\ref{table:primitive} shows how the derived operations fall into three categories and can address the aforementioned challenges.
%\ref{challenge:reasoning}--\ref{challenge:exposure-bias}.
First, as LLMs may have difficulty applying common sense reasoning or
complex inference to nuanced problems (C.1), the \emph{Classification} operation can act as a \textbf{validation check} or triage, before more steps are carried out (Table~\ref{table:primitive}a). 
For example, a chatbot may need to first classify the type of question a user is asking before providing adequate responses. 
Second, to alleviate exposure bias (C.2, the inability to generate long and diverse text), some operations can be used to \textbf{query small chunks of new content} (Table~\ref{table:primitive}b), so as to gradually build up the generation diversity and length.
Three ways to get new content include
\emph{querying facts}, \emph{generating hallucinations}, and \emph{ideating} lists of items.
For example, in the peer review rewriting scenario (Figure~\ref{fig:flow_review}B), the separate \emph{Ideation} step per problem prevents suggestions for one criticism from being influenced by the other two criticisms. Finally, because LLMs may struggle with certain input prompt types, \textbf{reorganizing the prompt} could be helpful when its original form is convoluted. 
\emph{Rewriting} and \emph{Compose points} transform input into more
parsable forms, \emph{Information Extraction} elicits concise
information (C.3), and \emph{Split points} splits
text into smaller and more manageable units (C.1)---all are summarized in Table~\ref{table:primitive}c.
As we will see in a case study (Section~\ref{ssec:viz-debug}), translating JSON-formatted specifications to natural language descriptions helps LLMs parse the embedded information.

Chaining and its operations also have some parallels to crowdsourcing workflows. 
However, whereas sub-tasks in crowdsourcing are assumed to be feasible for a human worker (reviewed in Section~\ref{ssec:relate-crowdsourcing}), LLMs are more restricted in terms of tasks they can perform reliably, and thus the primitive operations presented are more scoped and granular. 
For example, Kittur \etal~\cite{kittur2011crowdforge}'s Partition-Map-Reduce workflow uses \emph{Split} and \emph{Compose Points} operations (in Figure~\ref{fig:flow_review}B), but does not indicate specifically how to transform the text (\emph{Ideation}), though it also targets collaborative writing.
%, which seeks to decompose difficult tasks into smaller ones that are easier for human workers to accomplish; 
%conversely, our operations naturally support a variety of crowdsourcing workflows.
%For instance, \emph{Information Extraction} and \emph{Classification} are crucial components of the Find-Fix-Verify workflow for effective text editing~\cite{bernstein2010soylent} (the ``fix'' step may be \emph{Rewriting}, \emph{Compose points} or \emph{Generation}, depending on the goal). 
%Similarly \emph{Split} and \emph{Compose Points} would enable key steps in Divide-and-Conquer~\cite{kulkarni2011turkomatic}, and support parallel task patterns~\cite{little2010exploring}.
%however, while 
%In addition to 
%crowdsourcing workflows primarily emphasize how tasks are decomposed and re-composed (\ie how to arrive at sub-tasks and recombine results), the LLM operations presented above further scope what each sub-task should be.
%As mentioned in Section~\ref{ssec:relate-crowdsourcing}, this is essentially due to the different objectives of task decomposition:

%%%%%%%%%%%%%%%%%%%%%%%%%%%%%%%%%%%%%%%%%%%%%%%%%%%%%%%%%%%%%%%

\subsection{Designing Operations for LLM Chaining}
\label{ssec:primitive}

\begin{figure*}[t]
\centering
\includegraphics[trim={0 21cm 42.5cm 0cm}, clip, width=0.92\linewidth]{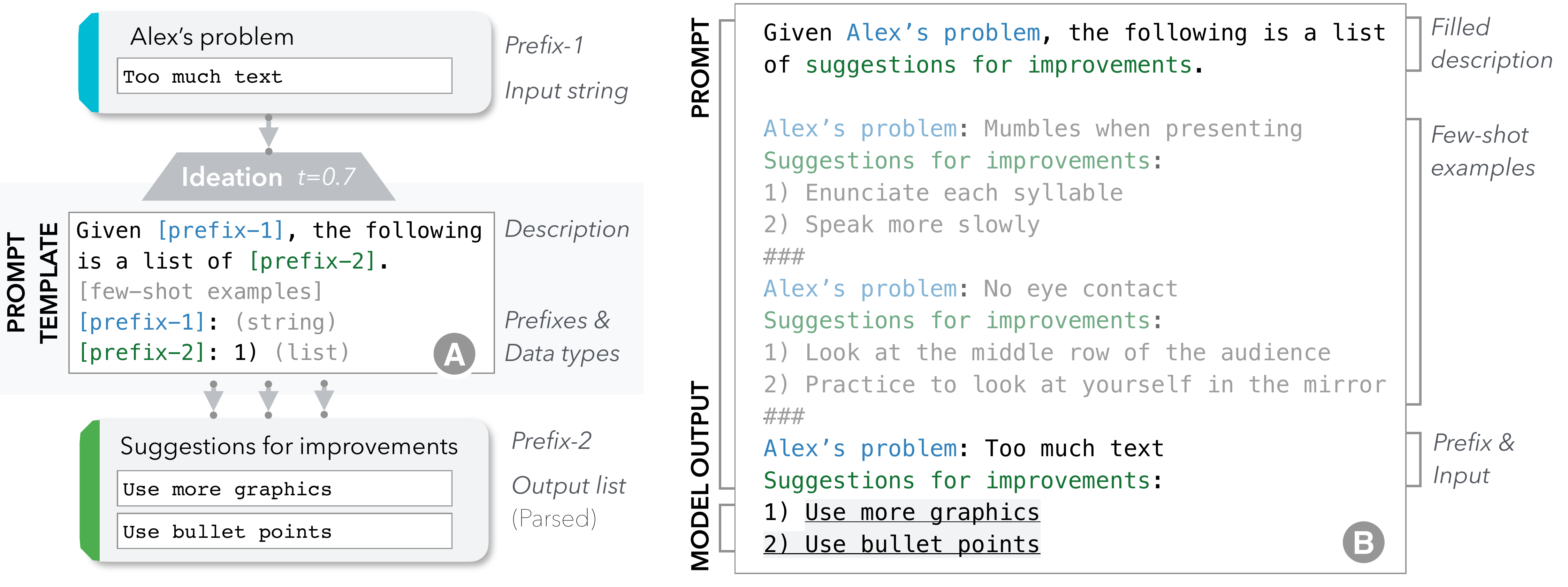}
\vspace{-5pt}
\caption{
An example of how to create an LLM step using a \textbf{prompt template} (A), using the \textit{Ideation} step of the peer review writing scenario (from Figure \ref{fig:flow_review}) as an example. For the peer review scenario, the \textit{Ideation} operation takes in a problem (\eg too much text) as input, and produces suggestions for improvement as output, but the prompt template allows the \textit{Ideation} operation to take in any custom inputs and outputs. %that ideates suggestions for presentation problems, with data layers and the default template for the \emph{Ideation} operation (A).
The template includes placeholders for the input ({\color{cplace1}{prefix-1}}), output ({\color{cplace2}{prefix-2}}), and (optional) few-shot examples.
%, as well as their corresponding \emph{data types} (\eg string or list).
%It also includes a default \emph{temperature} that controls the randomness of the LLM generation.
(B) shows the actual prompt after filling in the placeholders in the prompt template.%, while the \emph{Ideation} prompt in Table~\ref{table:primitive} represents the zero-shot version with just the task description.
}
\Description{The figure shows an example of how to create an LLM step using a prompt template. For the peer review scenario, the Ideation operation takes in a problem (e.g. too much text) as input, and produces suggestions for improvement as output, but the prompt template allows the Ideation operation to take in any custom inputs and outputs.}
\vspace{\anon{-10pt}{-5pt}}
\label{fig:primitive_param}
\end{figure*}

An LLM Chain consists of multiple \textbf{steps}. Each step is defined by an \textbf{LLM operation}, which takes in input data and produces output data (which we call \textbf{data layers}). 
For example, the \emph{Split point} operation in Figure~\ref{fig:flow_review} takes in the \colordot{coriginal}{initial feedback for Alex} as input, and produces a list of \colordot{cproblem}{presentation problems} (\exinline{too much text}, \exinline{no clear structure}, etc.) as output.
LLM Chains are constructed by connecting these steps through shared data layers.
%: the output of one operation becomes the input to the next.
In the same example above, the \emph{Ideation} operation comes after the \emph{Split points} operation, taking a (previously generated) \colordot{cproblem}{problem} as input and producing \colordot{csuggestion}{suggestions for improvements} as output.

Each step of an LLM (an operation and its data layers) is accomplished through a natural language \textbf{prompt}. While prompts are task-dependent, they can have some task-agnostic properties. For example, the prompt for the \emph{Classification} operation would likely contain the verb \exinline{classify}, regardless of what is being classified. These keywords help set an LLM operation's scope and expectations~\cite{o2021context}.
We aim to abstract these task-agnostic properties into default parameters for each operation (Figure~\ref{fig:primitive_param}A), so as to provide consistent starting points for interacting with LLM Chains across use cases.
Using the \emph{Ideation} operation as an example, we show how we design these parameters to satisfy the following three requirements for chaining, and how they help to build the \emph{Ideation} prompt shown in Table~\ref{table:primitive} and Figure~\ref{fig:primitive_param}B.

\textbf{Operations need to invoke the desired functionalities}, through prompt design.
%As mentioned in Section~\ref{ssec:relate-llm}, LLMs can be customized to produce specific functionalityare capable of in-context learning through natural languagetextual \emph{prompts}.
\todo{Clarify the descriptions for zero-shot and few-shot prompts. Examples are not enough; there were no definitions or descriptions.}
To date, the most common patterns for prompting are either \emph{zero-shot} or \emph{few-shot} prompts\added{, depending on how many demonstrating examples are provided in the prompt}~\cite{brown2020language}. 
\highlight{Zero-shot prompts directly describe what ought to happen in a task}: \eg we can enact \emph{Ideation} with a \emph{task description} prompt \exinline{Given Alex's presentation problem, the following is a list of suggestions.}
In contrast, \highlight{\emph{few-shot} prompts show the LLM what pattern to follow by feeding it examples of the desired input and output data}: \exinline{Problem: mumbles when presenting, Suggestion: enunciate each syllable, Problem: too much text, Suggestion:} (full prompt in Figure~\ref{fig:primitive_param}B). 
Given these prompts, the LLM might produce a reasonable suggestion, \eg \exinline{use more graphics on the slides.} Zero-shot prompts can also be easily transformed into few-shot prompts, by appending examples to the initial zero-shot task description.
In either case, prompts commonly include meaningful names as \textbf{prefixes} (\exinline{Problem:} and \exinline{Suggestion:}) to demarcate structure, which helps re-emphasize the desired intent~\cite{vaswani2017attention}.
Following this convention, we build our prompts to include \emph{task descriptions} followed by \emph{prefixes}.
%(\exinline{Alex's problem: too much text, Suggestions for improvement:}). 
%However, other prompt designs could be implemented for chaining.
Aside from the prompt itself, we also associate with each LLM operation a default \textbf{temperature setting}: a model parameter that influences the randomness of the LLM generation.
For instance, creative operations like \emph{Ideation} benefit from a higher temperature ($t\text{=}0.7$) than  more factual or deterministic tasks like \emph{Classification} ($t\text{=}0.0$)~\cite{gpt3instruct}.

\textbf{Operations should be able to take custom data layers as inputs and outputs.}
Though our walkthrough example takes in \exinline{Alex's presentation problem} and generates \exinline{Suggestions}, in theory an operation should be able to handle any custom data layers. 
We thus create \textbf{prompt templates} to support a wide range of scenarios, with placeholders for input and output data.
%We show one possible implementation of a prompt template in Figure~\ref{fig:primitive_param}A: the placeholders in the prompt template takes in two types of information for each data layer: 
%(1) a \textbf{prefix} that names the desired input/output (\eg \primPrefix{cinput1}{Alex's problem} for the input layer), and 
%(2) a \textbf{detail} string that optionally explains additional information (\eg \primPrefix{cinput1}{Alex's presentation problem}).
The template allows us to build LLM steps simply by filling in the placeholders with definitions on data layers, as demonstrated in Figure~\ref{fig:primitive_param}.
%, as well as placeholders for few-shot examples.
In particular, we include key verbs and nouns~\cite{o2021context} in the template, to best reflect the operation objective (\eg \exinline{a list of} for \emph{Ideation}, \exinline{classify} for \emph{Classification}).
The template also accepts optional few-shot examples.
We can build the few-shot prompt in Figure~\ref{fig:primitive_param}B if we provide those pairs of problems and suggestions, or default to just the zero-shot version in Table~\ref{table:primitive} when examples are not readily available. 
Though we provide this as one example of a prompt template, we do not claim it to be exhaustive as there may be other equally effective ones.

\textbf{Operations should handle parsing of the expected input/ output data types.} Different data layers may take on different data types.
For example, the Split step (Figure~\ref{fig:flow_review}$b_1$) produces a \emph{list} of problems, but only a \emph{single} problem is the input to each subsequent Ideation step ($b_2$). 
To handle different formats in different steps, in each operation’s definition, we define the required data types per operation (e.g. ``list'' in Figure~\ref{fig:primitive_param} for \emph{Ideation}), along with the corresponding parsing necessary to produce the expected data type (\eg split each row of the numbered list into an item).
%In Figure~\ref{fig:primitive_param}, because \emph{Ideation} accepts a single string, and outputs \emph{a list of} suggestions, its prompt template includes a \texttt{``1)''} right before the output text, to encourage the generation of a list, and the output is also parsed into lists based on such numbering.

Empirically, we find these defaults to work reasonably well across domains (see later sections~\ref{sec:user-study} and~\ref{sec:case-study}). 
Still, we note that our defaults here are just one example of possible operation implementations; in our review of existing demos, there appeared to be many diverse prompting strategies even for the same task. 
%While end-user authoring of chains is beyond the scope of this paper, 
We hope the prompt templates provided here may serve as a starting point for Chain designers or users to modify. 
In the next section, we demonstrate how these designs serve as the underlying data structure for interactive Chain execution by end-users.

%Still, we note that our defaults here is just one example of possible operation implementations. 
%We designed them to follow the field conventions, and to create consistent structures operations, but these defaults are far from a complete scoping of the LLM prompt engineering space.
%Prompts curated per task are usually more effective~\cite{reynolds2021prompt}, and we observe many diverse prompting strategies for the same task throughout our surveys (\eg question answerings with chatbots, purely rely on descriptive prefixes, etc.)
%Therefore, we expect end users to customize and iterate on top of the defaults, or even create their own arbitrary prompts (also discussed in Section~\ref{sec:discussion}).

%When constructing a Chain, for example, they should define the data layers and select the operations; and if they wish to customize a step, they might refine the task description, or optionally include application-specific example tasks.
%In the next section, we demonstrate how these designs serve as the underlying data structure for interactive Chain execution and customization.

%%%% definition.tex ends here %%%%

%%%% ui.tex starts here %%%%

\section{Interactive User Interface}
\label{sec:ui}

%To enable users to interact with LLM Chains, we designed an interactive interface which supports users in executing Chains, modifying the prompts and intermediate model outputs for each step, and customizing the Chains.
We designed an interface that helps users execute and customize LLM Chains interactively.
%to support users interacting with LLM Chains, both in terms of executing them and customizing them.

\begin{figure*}[t]
\centering
\includegraphics[trim={0 8.8cm 36cm 0cm}, clip, width=1\linewidth]{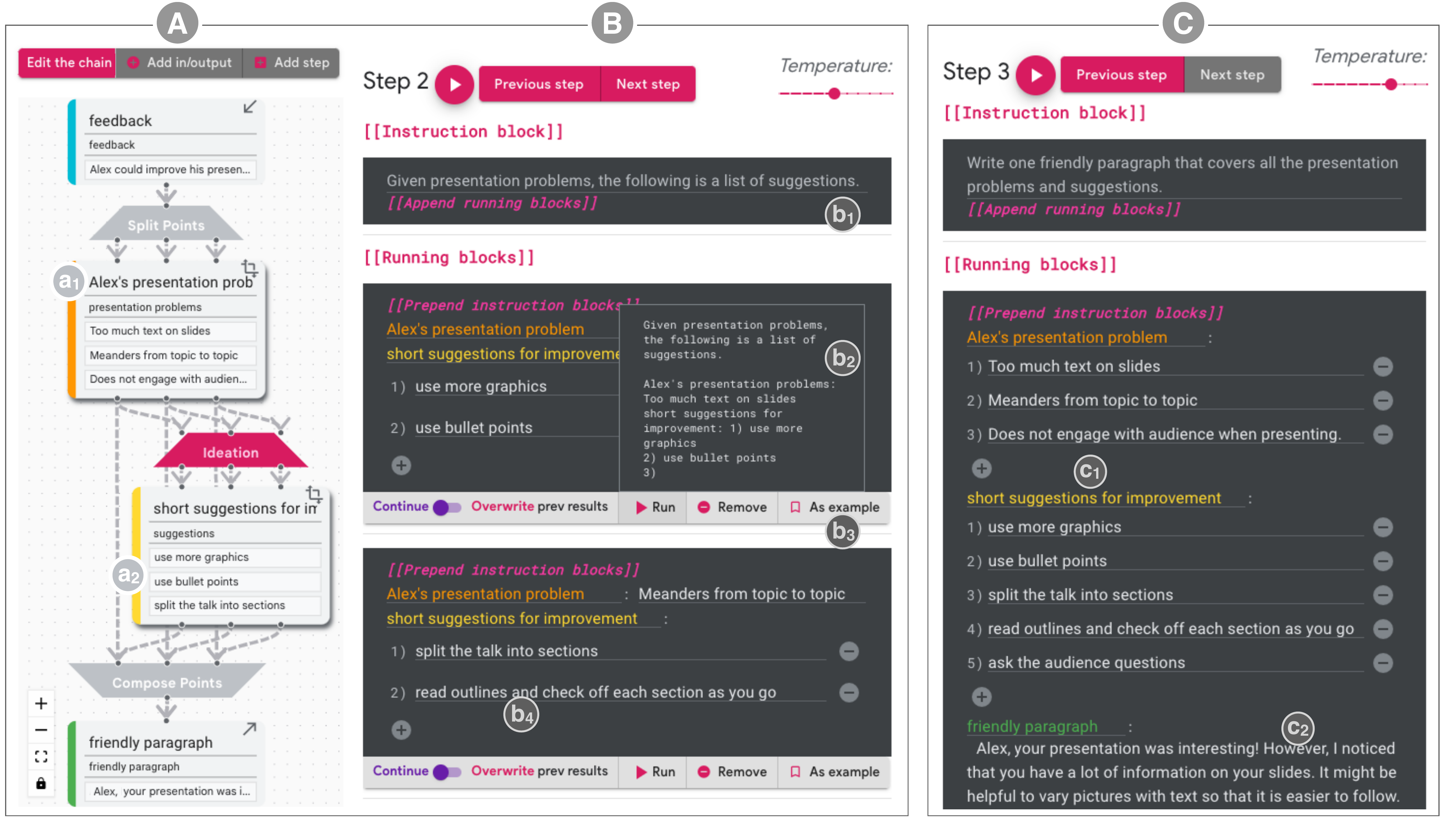}
\vspace{-15pt}
\caption{
An overview of the interface, reflecting the peer review rewriting example in Figure~\ref{fig:flow_review}.
It consists of 
(A) a \emph{Chain view} that depicts the high level Chaining structure, and 
(B/C) a \emph{Step view} that allows for refining and executing each LLM step.
The interface facilitates tracking the progress of the LLM Chain.
For example, when moving from step 2: \emph{Ideation} (B) to step 3: \emph{Compose Points} (C), the previously generated presentation problems and suggestions become inputs for the final paragraph.
A demonstration is available at \url{https://youtu.be/QFS-1EWlvMM}.
}
\Description{An overview of the interface, reflecting the peer review rewriting example in Figure~\ref{fig:flow_review}.
It consists of 
(A) a \emph{Chain view} that depicts the high level Chaining structure, and 
(B/C) a \emph{Step view} that allows for refining and executing each LLM step.
The interface facilitates tracking the progress of the LLM Chain.}
\vspace{\anon{-10pt}{-5pt}}
\label{fig:ui}
\end{figure*}

\subsection{Design Rationales}

\todo{Clarify which users are being referred to when describing Rationale R.2}
Over the course of several weeks, we designed and iterated on the prototype with feedback from \highlight{four pilot users }\added{(software engineers and designers who have experience designing LLM prompts)}, producing three design rationales for the final interface. 

\begin{enumerate}%[labelwidth=*,leftmargin=1.8em,align=left,label=R.\arabic*]
\item[R.1] %\label{design:visual-structure} 
\textbf{Visually reflect the underlying Chaining structure.}
In early prototypes, we explained the Chain structure using a static slide deck that highlighted the \emph{data} produced at each step (\eg problems, suggestions for improvement, and final paragraph in Figure~\ref{fig:flow_review}). In reaction, users expressed a desire to understand the \emph{operations} taken at each step to arrive at these data layers (split points, ideation, compose points), and wanted to visually track progress through the Chain.
To achieve this, we designed the interface to reflect not only the data layers, but also the LLM details within each step.

\item[R.2] %\label{design:control-level}  
\textbf{Provide controls at different granularities.}
Pilot users favored flexible controls. We observed users frequently making local fixes on intermediate data points that flow between LLM steps, and therefore designed the UI to allow in-place editing, without explicitly requiring a switch to editing mode.
Some users also voiced an interest in iterating on alternative Chaining structures (\emph{``Can I change this step with...''}).
We therefore conclude that the interface should support modification of LLM Chains both locally (\eg changing one task description or intermediate model output) and globally (\eg changing how the steps are connected). Because global changes have more impactful consequences (they may overwrite the underlying Chain structure), we designed the UI to require a switch to editing mode for this type of changes.
%, but not for local edits which seemed to occur much more frequently.
%Further, the ease of interaction should in line with the level of control.
%We observed users frequently making local fixes on intermediate data points that flow between LLM steps, and therefore design the UI to allow in-place editing, without explicitly switching to editing mode.
%Rewiring the entire Chain, on the other hand, had more impactful consequences and tended to overwrite the underlying structure, so we hide this level of modification by default.

\item[R.3] %\label{design:mimic-nl} 
\textbf{The structured controls should still reflect the natural language interaction supported by LLMs.}
In an early prototype, we formatted the data as structured tables with each data layer being a column, but received feedback from two users that making text edits in cells felt unnatural as they lost the sense of interacting with the model through natural language.
To retain a natural interaction experience, we keep these structures as in-line text fields.
\end{enumerate}

\subsection{Interface Design and Implementation}
We design the interface in Figure~\ref{fig:ui} following these design rationales above, which consists of two primary views: the \emph{Chain view} (Figure~\ref{fig:ui}A), and the \emph{Step view} (Figure~\ref{fig:ui}B/C).

The \textbf{Chain view} (Figure~\ref{fig:ui}A) depicts the high level Chaining structure through a flow chart.
%\paragraph{Visual flow view} 
%Figure~\ref{fig:ui}A depicts the high level Chaining structure through a flow chart.
%\footnote{The implementation is based on the open-source React Flow library: \url{https://reactflow.dev/examples/}.}
It contains three primary visual cues that closely reflect the
underlying design (R.1) described in Section~\ref{ssec:primitive}.
First, we use grey glyphs to represent \emph{LLM operations}, with shapes indicating 1-1 (rectangle, for operations like \emph{Rewriting} in Table~\ref{table:primitive}), 1-N (trapezoid, \eg \emph{Ideation} operation), and N-1 data mappings (inverted trapezoid, \eg \emph{Compose points} operation).
Clicking on these glyphs allows users to choose which step to zoom into (highlighted in pink), and the Step view would change in response.
Then, we use rectangles with colored stripes to represent \emph{data layers}.
%, and include prefixes and details in the rectangle.
\todo{Clarify if the number of arrows matter}
Users can preview their data entries through white rows (\eg Figure~\ref{fig:ui}$a_1$ and $a_2$), which are updated after each LLM execution, and thus track Chain execution progress.
Finally, we link these elements with dotted-line arrows to highlight which data output serves as the input to which step, and \highlight{use the number of arrows going out of an operation to re-emphasize the data mappings (\eg multiple \colordot{cproblem}{problems} coming out from \emph{Split points}, \added{which is approximated with three lines}, and a single \colordot{crewrite}{paragraph} out of \emph{Compose points}).}
%Finally, we link these elements with dotted-line arrows to highlight which data output serves as the input to which step, and use the number of arrows to roughly reflect whether the step expects a single input/output (\eg a single \colordot{coriginal}{feedback} paragraph going into \emph{Split points}), or multiple of them (\eg multiple parallel \colordot{cproblem}{problems} going into \emph{Ideation}).

%, which would affect the number of parallel paths we see in the within-step view below.

On the right, the \textbf{Step view} (Figure~\ref{fig:ui}B) allows users to explore each LLM step by interacting with inputs, outputs, and the underlying prompt structure.
%\paragraph{Within-step view} On the right, this view in Figure~\ref{fig:ui}B allows users to explore each LLM step by interacting with inputs, outputs, and the underlying prompt structure.
It is divided into an \emph{instruction block} and several \emph{running blocks} to handle \emph{parallel paths}.
Each of these parallel paths translates to a different LLM invocation; they share some common parts in their prompt strings, while having other parts being distinct from each other.
We use the running blocks to hold the unique parts, and the instruction block to hold the shared sub-string is pre-pended to \emph{all} running blocks, such that they are combined to form the full prompt.
%Parallel paths need to be handled carefully because they may share some parts of a prompt string in common (held in the instruction block), but have others parts of a prompt that are distinct from each other (held in the running blocks).
%The running blocks represent parallel paths, each denoting a different LLM invocation, and the instruction block is the shared part of the prompt that is pre-pended to \emph{all} running blocks.
%They are combined to form the full prompt for that step.
For example, Figure~\ref{fig:ui}$b_2$ is the final prompt for the step that generations suggestions for the problem \exinline{too much text.}
It starts with the content from the instruction block ($b_1$), and merges the text in the running block thereafter, ignoring the other parallel running blocks.

Every running block visually resembles a textarea with a number of editable text fields.
It shows the prefix fields before colons (\eg \colordot{csuggestion}{Short suggestions for improvement}, $c_1$) in the same color as the data layer rectangles, which helps users distinguish between data layers. It also includes text fields ($b_4$, $c_2$) for the model output for that step. The number of text fields (\eg 1 vs. N) are consistent with the data types defined for the primitive operation for that step. 
%Thereafter, it switches between a text field ($c_2$) and a list of text fields ($b_4$) to reflect whether the data type of the operation is strings, or a list of strings.
%All these fields are editable, with the former affecting data layers, and the latter affecting intermediate data entries.

This view also handles the per-step execution.
Users can click the small ``run'' button to execute each running block individually. Alternatively, users can use the Play button on the top to run all the parallel blocks at once and compare their results.
To improve natural language interaction transparency (R.3), running a block also triggers a preview of the final prompt text ($b_2$).
The output is then parsed and added to the corresponding field ($b_4$, $c_2$) for users to further iterate on.
%Finished running blocks can be used ``As example'' ($b_3$), \ie moved into the instruction block and used as few-shot examples for other remaining running blocks, thereby gradually improving the quality of the instruction block.

\textbf{Interactions and controls.}
Notably, there are three levels of control available with this
interface (R.2), from local customization of prompts to global modification of the LLM Chain structure, each with clear cues on its impact.
First, users can customize the prompt for a particular step, \eg by changing its task descriptions.
Since the customization only applies to the current step, all other views remain unchanged. 
Second, users can customize the model output for that step by adding, deleting, or editing content (\eg editing \exinline{read outlines} to \emph{emphasize main points} in $b_4$), or rename data layers (\eg rephrasing \exinline{Alex's presentation problems} as \exinline{Criticisms of Alex} in $a_1$).
These changes impact both the current step in focus as well as other steps involving the shared data layers (\eg \emph{Compose Points} takes in both the \exinline{problems} and the \exinline{suggestion} layer), and thus they can be changed either in the colored rectangles in the Chain view, or through text fields in the Step view.
Finally, users can more aggressively modify the Chaining structure
itself by adding, removing and rewiring operations or data layers in
the Chain view through intuitive visual programming (R.3).
The change would then cause the entire Chain to re-render, with the defaults (\eg temperature, instructions) refreshed.
%While alternatively we can create a separate view for defining data layers and LLM steps, 
%We choose to allow direct edits on the visual flow, so 
%Because the re-rendering would corrupt the underlying data mapping, 
%Note that unlike all the previous customization which are enabled by default, the Chaining reconstruction is only accessible if users explicitly click on ``edit Chain.'' ($a_1$) 
%The within-step view is disabled and hidden accordingly, to convey the message that the existing stages would be overwritten once the edit is complete.

%%%% ui.tex ends here %%%%

%%%% eval_user.tex starts here %%%%

\section{User Study}
\label{sec:user-study}

To understand how Chaining affects the user experience of accomplishing tasks with LLMs, we conducted a within-subject user study comparing Chaining with a state-of-the-art baseline interface, on two user tasks.

\subsection{Study Design}

\revised{\textbf{Underlying LLM.} 
All of our experiments (including our baseline interface introduced below) and \emph{each step of the Chaining interface} rely on exactly the same underlying LLM: LaMDA~\cite{thoppilan2022lamda}\footnote{We used a non-dialog version of the model.}, a 137 billion parameter, general-purpose language model.
This model is roughly equivalent to the GPT-3 model in terms of size and capability: it is trained with more than 1.5T words of text data, in an auto-regressive manner using a decoder-only Transformer structure which is useful for text generation.
It has comparable performances with GPT-3 on a variety of tasks, and behaves similarly in its ability to follow prompts.
Note that we only use this model to represent the recent class of LLMs; essentially, the chaining interface is model agnostic, and is compatible with any LLM that has in-context learning capability.}

\textbf{Systems.}
We compared \chain with \sandbox, an interface that looks aesthetically similar to the Chaining interface, but without the Chaining functionality.
We based the \sandbox interaction on GPT-3 playground,\footnote{\url{https://gpt3demo.com/apps/openai-gpt-3-playground}} the standard online interface for LLMs. 
%Like GPT-3 playground, \sandbox 
It presents a single textbox with a run button, which allows the user to enter the text prompt, run the model on that prompt, and then view the model result in the same textbox, with the ability to edit that result and then continue to iterate.
\highlight{\removed{Both \chain and \sandbox use the same underlying LLM, a 1XXB billion parameter, internal language model
%\footnote{\anon{The model name is anonymized for review.}{The model is a version~\cite{lambda_blog} of the Meena model from Adiwardana \etal~\cite{adiwardana2020towards}.}}
that behaves in a very similar way to GPT-3 in its ability to follow prompts.}}
Like the \chain interface, the \sandbox also allows users to adjust the temperature setting through a knob.

\begin{figure*}[t]
\centering
\includegraphics[trim={0 28.6cm 35cm 0cm}, clip, width=1\linewidth]{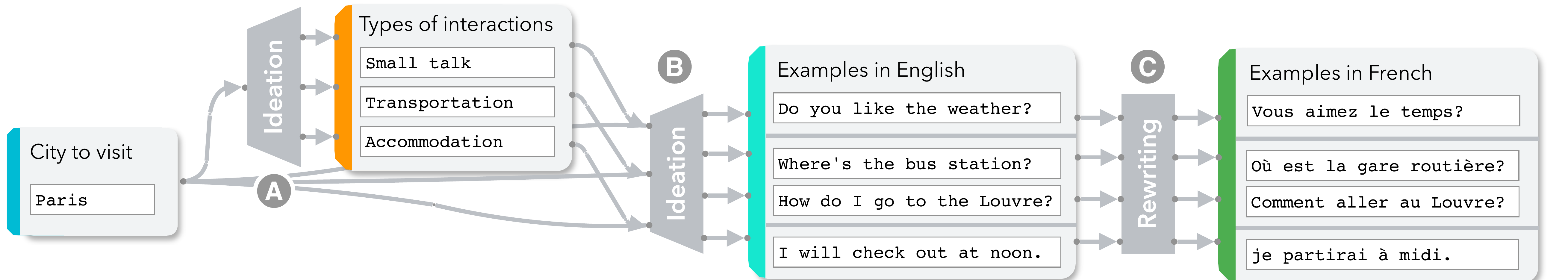}
\vspace{-15pt}
\Description{The LLM Chain for flashcard creation.}
\caption{
The LLM Chain for flashcard creation, with:
(A) An \textbf{Ideation} step that brainstorms the \colordot{cinteraction}{types of interactions} that we might encounter when \colordot{ccity}{visiting a given city} (Paris), 
(B) Another \textbf{Ideation} step that creates a list of \colordot{cen}{English examples} for each \colordot{cinteraction}{interaction type}, and
%(split by grey dividers)
(C) A \textbf{Rewriting} step that translates each \colordot{cen}{English example} into \colordot{cfr}{French}.
}
\vspace{\anon{-10pt}{-5pt}}
\label{fig:flow_flashcard}
\end{figure*}

\textbf{Tasks.}
We conducted the study using two tasks: 
peer review writing, and personalized flashcard creation, as they reflect different types of challenges (as explained below), and are both commonly used in user-centered task scenarios~\cite{cai2014wait, edge2011micromandarin, cambre2018juxtapeer}.
In the peer review writing task (``\textbf{\taskReview},'' our walk-through scenario),
the user is given a paragraph (the same as in Figure~\ref{fig:flow_review}) outlining three different problems in an imaginary person's presentation style, 
and their task is to write a friendly paragraph with 1-3 suggestions for each problem. 
% (too much text on slides, meanders from topic to topic, etc.)
In flashcard creation (``\textbf{\taskFlashcard}''), participants were asked to create at least ten English-French sentence pairs they could use while traveling in Paris, and to make them as diverse as possible while being personalized to their own travel goals.
%\footnote{Though French/Paris was the default setting, we allowed participants to freely choose alternative languages and destinations.}
%Details are in Appendix~\ref{sec:appendix-user-study}.

Though both tasks are possible when using an LLM without any LLM Chains, they present different types of challenges which could potentially be improved through Chaining.
The \taskReview task implicitly involves multi-step reasoning
(Challenge C.1 in Section~\ref{sec:pipeline}): 
to create a thorough and constructive review, one needs to identify each problem, provide suggestions per problem, and compose all the suggestions into one paragraph.
The \taskFlashcard task, on the other hand, exposes the challenge of
having sufficient diversity in light of LLM exposure bias (C.2).
In the \chain condition, we built a default Chain for each task. 
The Chain for \taskReview in Figure~\ref{fig:flow_review} reflects the three aforementioned steps (as explained before);
%For example, the Chain for \taskReview involves a \emph{Split} operation to ensure that each problem is addressed, and separates the sub-task of generating suggestions from the sub-task of composing a constructive paragraph; 
the Chain for \taskFlashcard (see Figure~\ref{fig:flow_flashcard}) sources additional content from the \modelname like \colordot{cinteraction}{types of interactions} in a trip, which can help the user diversify the flashcards.

\textbf{Study procedure.}
Before the study, participants completed a 30-minute tutorial that summarized the concept of LLMs and demonstrated how both \sandbox and \chain work.\footnote{We took inspiration from the OpenAI: \url{https://beta.openai.com/docs/introduction/prompt-design-101}; the task used for tutorial is in Appendix~\ref{ssec:appendix-user-study-metaphor}.}
\added{They were told upfront that both systems rely on the same underlying LLM.}
Then, in an hour-long study, participants performed a randomly selected task (\taskFlashcard or \taskReview), once with each interface (\sandbox and \chain), whose orders were counterbalanced.
We first briefed participants on the task, and then asked them to accomplish it %in conjunction 
with \modelname's help in each interface until they were satisfied with the final results, or until they reached 25 minutes.
Since LLM Chains came with automatically generated prompts (by filling in the templates), we similarly offered several default prompts for \sandbox that we knew to work reasonably, so that both interfaces had a fair starting point for prompt engineering (detailed in Appendix~\ref{sec:appendix-user-study}).
We encouraged participants to think aloud and describe their actions as they completed the task.

In the \chain condition, participants were asked to first stick to the default Chain so that we could make consistent observations across participants in terms of how they \emph{use} Chains. 
In the process, they could modify any other aspect (\eg the prompt, the intermediate model outputs, etc.)
At the end, we gave participants the option to \emph{modify} the default Chain, so that we could observe how they would expect the \modelname to assist them beyond the default design.
Finally, participants completed an exit survey and a semi-structured interview.
They rated their experience using each interface along various dimensions.
\removed{, including transparency, controllability, collaboration, support for their thought process, and quality of the final result.}
\todo{Clarify how the dependent variables were selected for the survey of self-perceived experience.}
\added{
These dimensions were chosen to reflect the effectiveness of the human-AI collaboration (\eg support for their thought process, quality of the final result), and core user-centered challenges in human-AI systems~\cite{amershi2019guidelines, buschek2021nine,huang2020ai} (\eg transparency, controllability, and sense of collaboration).
}
They also verbally compared their impressions of the two interfaces, and envisioned possible use cases for them.

\textbf{Collected data.}
We collected and analyzed three sets of data.
\todo{Clarify your Likert scale: What do the different points of the scale stand for?}
First, to assess participants' \textbf{self-perceived experience}, we used a standard seven-point Likert Scale~\cite{likert1932technique} to collect all ratings from the exit survey\added{, with one being ``Strongly disagree'' and seven being ``Strongly agree'' with the statement in question (\eg for system Transparency: \exinline{\added{The system is transparent about how it arrives at its final result}}). 
Detailed survey questions are listed in Appendix~\ref{ssec:appendix-user-study-survey}.}

We also observed and recorded their entire task completion sessions, and later transcribed their comments and experience for qualitative analysis.
Second, to quantify their \textbf{interaction mechanisms and behaviors}, we logged their interactions with the two interfaces. 
We were particularly interested in how participants reacted and iterated on model outputs, so we sorted their interactions with text fields by:
(1) whether participants mainly relied on running the model again to get a different result (\texttt{Consecutive run}), or if they also edited the prompt in between (\texttt{Edited});
and (2) when they edited the prompt, how dependent it was on the existing model generation: whether they closely \texttt{CURATED} and refined the model outputs, loosely interacted around them by \texttt{CREATING} completely new content, or tried again by \texttt{UNDOING} the outputs.
The detailed categorization criteria is in Appendix~\ref{ssec:appendix-user-study-clickstream}.
Third, to assess the \textbf{task outcome}, we logged the final reviews and flashcards participants created.
%Two non-participants, blinded to the condition, independently performed anonymized, paired comparisons on the results from \sandbox and \chain from each participant, by selecting the result that better satisfied the corresponding task goal.
Blinded to the condition, two non-participants performed anonymous, paired comparisons on results from each participant in \sandbox and \chain, choosing the result that satisfied the task goals the best.

\textbf{Participants.}
\todo{For example, provide mean and stdev of participants’ age. Also clarify how were participants recruited.}
We recruited 20 participants \added{using email lists that reach a wide range of practitioners (\eg UX designers, linguists, data analysts) at a large software company. Eight participants were 26-35 years old, eight aged 36-45, two aged 46-55, one 56-65, and one 18–26.}
As there is an initial learning curve associated with LLM capability, we required that participants had at least seen an LLM example before.
%, but did not require all participants to have prompting experience. 
Among those we recruited, half of the participants had no prompting experience but had seen online demos powered by LLM models, whereas the other half had some basic experience using default text prompts. 
Further, as the goal of Chaining is to use LLMs to assist with human tasks, we sought to recruit potential \textit{users} of ML/LLM who would benefit from interacting with the models, rather than ML model experts or \textit{creators}. 
Thus, our participants included technically knowledgeable but non-ML software engineers, linguists, UX designers, and data analysts who worked in a wide range of domains (\eg health, privacy, cloud storage, etc.).
Each participant spent approximately 90 minutes total in our study, and received a \$40 gift certificate for their time.

\subsection{Quantitative Results: Increased Transparency \& Control, and Higher-quality Task Outcome}

%%%%%%%%%%%%%%%%%%%%%%%%%%%%%%%%%%%%%%%%%%%%%%%%%%%%%%%%%%%%
\begin{figure*}[t]
\centering
\includegraphics[trim={0 2cm 0cm 0cm}, clip,width=0.7\linewidth]{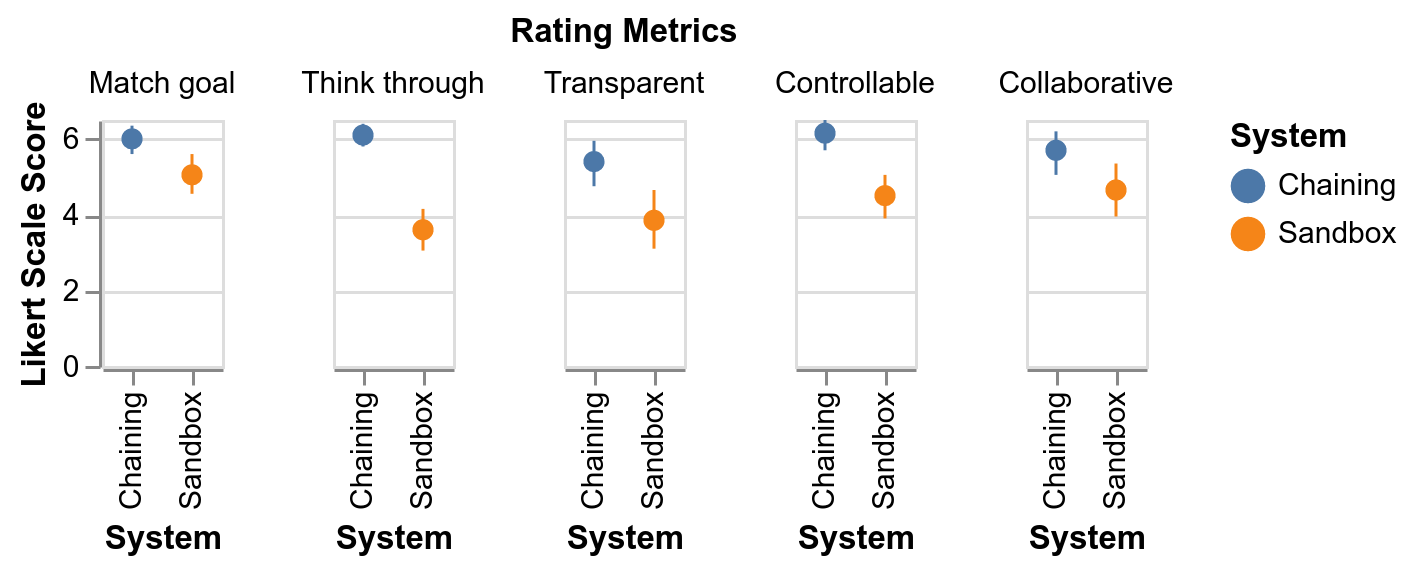}
\vspace{-5pt}
\Description{Participants' ratings in the form of seven-point Likert scale questions, with 95\% confidence intervals. 
Using \chain, participants felt they produced results that better \emph{matched the task goals}, and that the system helped them \emph{think through} the task.
They also found \chain more \emph{transparent}, \emph{controllable}, and \emph{collaborative}.}
\caption{
Participants' ratings in the form of seven-point Likert scale questions (details in Appendix~\ref{ssec:appendix-user-study-survey}), with 95\% confidence intervals. 
Using \chain, participants felt they produced results that better \emph{matched the task goals}, and that the system helped them \emph{think through} the task.
They also found \chain more \emph{transparent}, \emph{controllable}, and \emph{collaborative}.
}
\vspace{-5pt}
\label{fig:likert}
\end{figure*}

\begin{figure*}[t]
\centering
\includegraphics[trim={0 32cm 32.5cm 0cm}, clip, width=0.95\linewidth]{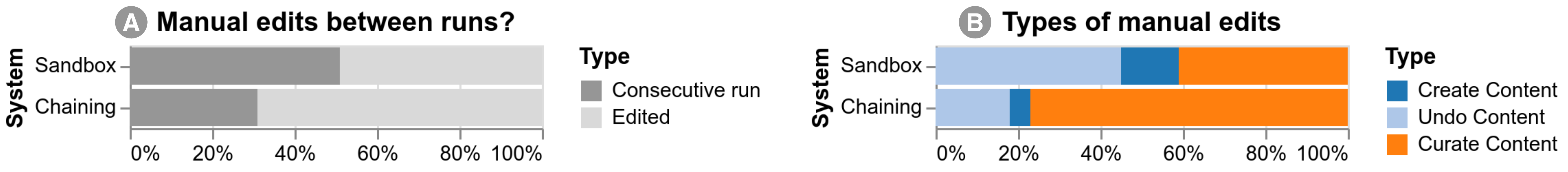}
\vspace{-5pt}
\Description{Distribution (based on the logged interactions) of how participants interacted with the prompts and model outputs, with and without chaining.
(A) They made more \emph{edits} in \chain (compared to just repeatedly running the model), and 
(B) They tended to \emph{curate} model outputs, rather than either deleting (\emph{undoing}) them entirely or manually \emph{creating} new content. 
}
\caption{
Distribution (based on the logged interactions) of how participants interacted with the prompts and model outputs, with and without chaining.
(A) They made more \emph{edits} in \chain (compared to just repeatedly running the model), and 
(B) They tended to \emph{curate} model outputs, rather than either deleting (\emph{undoing}) them entirely or manually \emph{creating} new content. 
}
\vspace{-5pt}
\label{fig:log_distribute}
\end{figure*}

%%%%%%%%%%%%%%%%%%%%%%%%%%%%%%%%%%%%%%%%%%%%%%%%%%%%%%%%%%%%%%%%%%%%%%%%%%%%%%%%
\todo{Report the task time differences for the different methods.}
\added{All the participants were able to complete the tasks in both systems within the given time: they spent $12.4 \pm 4.0$ minutes in \sandbox, and $14.6 \pm 5.4$ in \chain. Student's t-test did not show any significant difference between their completion time ($t=-.1.1, p=.278$).}
In analyzing subjective ratings from participants, the logged clickstreams, as well as the final generated results, we found:

First, \emph{\chain led to improved user experience in human-AI interactions.}
\added{We performed the non-parametric Wilcoxon signed-rank test to compare users' nominal Likert Scale ratings and,}
%\added{We perform the Likert Scale ratings fall into seven nominal categories and therefore does not follow normal distribution,  nominal We performed the non-parametric Wilcoxon signed-rank test to compare users' ratings:}
as shown in Figure~\ref{fig:likert}, participants felt that \chain helped them \textbf{think through the task better} (\chain $6.0 \pm 1.4$ vs. \sandbox $3.6 \pm 1.3$, $z=0, p<.001$), and \textbf{gave them more control} ($6.2 \pm 0.9$ vs. $4.5 \pm 1.3$, $z=3.0, p<.001$).
They also rated \chain as being more \textbf{collaborative} ($5.7 \pm 1.3$ vs. $=4.6 \pm 1.6$ , $z=25, p=.04$) and 
\textbf{transparent} ($5.4 \pm 1.3$ vs. $3.8 \pm 1.8$, $z=9.0, p=.002$).

Second, \emph{\chain shifted the types of edits participants made while interacting with the LLM.}
In \chain, participants were more likely to make manual interventions, whereas in \sandbox, they often re-ran the model (without changing the prompt) --- akin to ``rolling the dice again" in an attempt to get better output.
As shown in Figure~\ref{fig:log_distribute}A, this tendency to perform consecutive runs without altering anything from the previous run occurred $51\%$ of the time on average in \sandbox and $36\%$ in \chain.
Student's t-test shows the difference is significant: $t=3.5, p=.001$.\footnote{\added{The clickstreams fall into the continuous range of 0\%--100\%, and follows a normal distribution according to a D'Agostino-Pearson Test (\eg $p = 0.58$ for the ratio of consecutive runs).}}
\todo{Clarify the choice of Wilcoxon (non-parametric) and Student's T (parametric)  tests, based on the distribution of the data.}
%\footnote{\added{Note that we used the non-parametric Wilcoxon signed-rank test on the Likert Scale since the ratings are essentially nominal and does not follow normal distributions. On the other hand, the clickstream distribution has a continuous scale on 0\%--100\% and follows normal distribution according to D'Agostino-Pearson Test, and therefore fits Student's t-test.}}

The manual edits made were also finer-grained in \chain than in \sandbox (Figure~\ref{fig:log_distribute}B).
In \sandbox, people largely focused on either completely \texttt{UNDO} output and rerunning the model (45\% of the time on average), or manually \texttt{CREATING} their own content as input to the model (14\%). They only \texttt{CURATED} or modified existing text 41\% of the time.
On the other hand, in \chain people performed \texttt{CURATION} 77\% of the time, only doing \texttt{UNDO} and \texttt{CREATE} 18\% and 5\% of the time, respectively.
The shift to \texttt{CURATION} is significant, according to Student's t-test ($t=-6.75, p<.001$).

As a result, \emph{\chain led to higher-quality generations that met the task goal.}
The two independent raters consistently preferred \chain results $85\%$ and $80\%$ of the time, respectively.
The results also matched participants' own judgements in Figure~\ref{fig:likert} (see \emph{Match goal}) --- they preferred their own final results from \chain ($6.0 \pm 0.9$) to the \sandbox results ($5.0 \pm 1.1$, Wilcoxon signed-rank test, $z=11.0, p\textbf{=}.002$).

Aside from \emph{using} \chain, many participants were also able to iterate on and customize the underlying Chaining structure.
While five of them preferred the default Chains provided and didn't want to change them, the remaining 15 people were able to identify parts they found lacking and suggested at least one change.
11 of them successfully implemented and executed one of their own solutions.

%\textbf{collaborative} (\sandbox $=4.6 \pm 1.6$ vs. \chain $=5.7 \pm 1.3$, Wilcoxon $z=25, p=0.04$)
% \textbf{transparent} (\sandbox $=3.8 \pm 1.8$ vs. \chain $=5.4 \pm 1.3$, Wilcoxon $z=9.0, p=0.002$)
% \textbf{better results} (\sandbox $=5.0 \pm 1.1$ vs. \chain $=6.0 \pm 0.9$, Wilcoxon $z=11, p=0.002$)
% \textbf{finer-grained controls} \sandbox $=4.5 \pm 1.3$ vs. \chain $=6.2 \pm 0.9$, Wilcoxon $z=3.0, p<0.001$
% \textbf{think about task}: (Likert Scale $3.6 \pm 1.3$ vs. $6.0 \pm 1.4$, Wilcoxon signed ranks test $z=0, p<0.001$)

%%%%%%%%%%%%%%%%%%%%%%%%%%%%%%%%%%%%%%%%%%%%%%%%
\label{ssec:user-study-quan}

\subsection{Qualitative results: Chaining as Guardrails and Operation Manuals}
\label{ssec:user-study-qual}

%Our analyses of the transcribed comments shed further light on the reasons for the quantitative differences.
Through analyses of the transcribed think-aloud comments and semi-structured interviews, we further unpack the reasons behind the quantitative differences.
\todo{Clarify the choices for analyzing the qualitative results. For example, was a thematic analysis conducted when selecting participants' quotations?}
\added{Since we asked participants to explain their Likert Scale ratings, their interview responses naturally map to dimensions in Figure~\ref{fig:likert} like transparency, collaboration, etc. One author further sorted their think-aloud comments into the categories. Three researchers then conducted thematic analysis, examining relationships between categories and iteratively converging on a set of higher-level themes.}
In general, Chaining helped support human-LLM interaction by serving as
(1) a guardrail that helped users stay on track towards the task goal (Section~\ref{sssec:control} and~\ref{sssec:think}); and 
(2) an ``operation manual'' that implicitly explained how to use LLMs for less obvious objectives, and that provided channels for users to intervene (Section~\ref{sssec:collaboration}, \ref{sssec:transparent} and \ref{sssec:debug}).
%(1) a manual that explains how to use LLMs for less obvious objectives and tells end users how to intervene in order to achieve the desired sub-goals.
In the following sections, we present key themes on how Chaining improved the human-AI experience, as well as some additional challenges brought on by Chaining.

%%%%%%%%%%%%%%%%%%%%%%%%%%%%%%%%%%%
\subsubsection{Chaining helped users more fully capitalize on the model's latent capabilities.}
% original name Surfacing different sub-tasks help diversity the forms and objectives of human-LLM collaborations.
\label{sssec:collaboration}
\mbox{}\\
In \sandbox, participants tended to use the LLM for a single purpose, under-utilizing the \removed{full model capability} \added{model's full potential in supporting various kinds of tasks}. Four out of ten people in the \taskFlashcard task only used the model as a translator in \sandbox, even though they were provided with default prompts that demonstrated how to generate English sentences using the model. 
%In contrast, the \chain allowed users to leverage the model in multiple ways: generating diverse topics, generating English sentences for each topic, and then translating to French.
\todo{Clarify what are the two steps in the two-step process of the Sandbox condition}
In the \taskReview task, even though nearly everyone (nine out of ten) used a \highlight{two-step process of generating suggestions prior to merging them into the full paragraph (see the two-step prompt template in Appendix~\ref{ssec:appendix-user-study-review})}, three people only relied on the LLM to generate suggestions, and then manually merged them into the paragraph themselves, without LLM input.

There may be two reasons for these behaviors. 
First, \sandbox naturally affords single-operation interactions. 
Given this, it is not surprising that users would gravitate toward using the model only for a part of the task that seemed most likely to yield promising results given the \emph{status-quo} applications of machine learning (\eg translation), overlooking others that may seem less likely to succeed (\eg merging text into a paragraph). 
Indeed, some participants were unaware of less obvious sub-tasks (P4: \quoteinline{this is just a simple translation task} in \taskFlashcard).
Second, the friction of juggling multiple sub-tasks in \sandbox deterred some users from doing so. 
%For example, two participants who got the \chain condition first in their study order became aware that the model could generate English sentences, in addition to translating sentences. 
%However, they found it difficult to juggle between sub-tasks in \sandbox. 
Even participants who became aware of the Chaining structure (from getting the \chain condition first in their study condition order) struggled to replicate it using a single prompt. 
For example, P2 attempted to tackle both sub-tasks (generating diverse English sentences, and translating to French) simultaneously with a single prompt instruction: \exinline{Given the previous English sentence, translate it to French. Generate further English sentences relevant to travel in Paris.}
However, because the instruction was too nuanced for the model to follow, they eventually resorted to manually creating their own English sentences.

Ultimately, this inability to fully utilize the model led to lower quality final results in \sandbox.
For example, the flashcards had less topical diversity (P4: \quoteinline{I had limited diversity myself}) because the \emph{Ideation} step in Figure~\ref{fig:flow_flashcard}A was rarely ever leveraged. 
As a byproduct of the inadequate support, participants also found collaboration in \sandbox to be shallow (P5: \quoteinline{I'm doing all the specific work [creating English sentences] and it's just doing its one thing [translation]}).
In contrast, \chain allowed users to leverage the model in multiple ways.
%: generating diverse topics, generating English sentences for each topic, and then translating to French.
Seven participants particularly liked that they could accomplish multiple goals through the Chain, \ie acquiring model-powered diversity in the \emph{Ideation} step, while maintaining translation correctness in the \emph{Rewriting} step. 
This additional support may have contributed to participants shifting from creation (manually creating text from scratch) to curation (modifying model outputs) as shown in Quantitative Results (Figure~\ref{fig:log_distribute}B). Quoting P5, \quoteinline{I didn't need to give it as much, but it was giving me a lot.}
%\quoteinline{It did the brainstorming for me, and I only needed to pick out the things that worked for me, and it seemed like it had a lot of ideas.}
%It was really helpful to be able to prune the output or add more things. In some ways it felt like I didn’t need to give it as much, but it was giving me a lot of stuff. I did a lot more curation; In some ways it did the brainstorming for me, and I only needed to pick out the things that worked for me, and it seemed like it had a lot of ideas.

LLMs' diverse primitive operations and capabilities also led participants to consider other ways the model might be helpful.
For example, when asked to modify the Chaining structure itself, P1 in \taskFlashcard swapped the \emph{Ideation} step (which generated \colordot{cinteraction}{types of interactions}) with a \emph{Generation} step to produce \colordot{cinteraction}{a journal of my one day trip}, so the model could \quoteinline{think about what conversations can happen across my day trip} and provide \quoteinline{less generic context suggestions.}
The operations became inspirational here.
P12 and P20 in \taskReview both added a \emph{Classification} step to determine if the paragraph is in the right voice or if a suggestion is actionable, only once they realized the classification operation existed.

%%%%%%%%%%%%%%%%%%%%%%%%%%%%%%%%%%%
\subsubsection{The ability to isolate interventions and save progress enhanced controllability of LLM}
\label{sssec:control}
\mbox{}\\
Because each step of a Chain involves a separate run of the model, Chaining allowed users to control certain aspects of each sub-task \emph{independent of} others.
Four \taskFlashcard participants in \chain noticed that the desired model randomness should vary per sub-task, and tuned the temperature settings accordingly: they increased the temperatures in \emph{Ideation} steps to broaden the diversity and creativity of model responses (Figure~\ref{fig:flow_flashcard}A and B), and lowered it for \textit{Rewriting} to increase the chances of getting correct model output (Figure~\ref{fig:flow_flashcard}C). 
However, none of them did so in the \sandbox condition (\eg P5: \quoteinline{I realized my temperature was always high in sandbox. I should have had it low at translation, and high when I ask the model for English sentences.})
Many \taskReview participants also liked iterating on each of the presentation problems individually (\eg \exinline{To much text on slides} vs. \exinline{No clear structure}) without affecting the others.

This well-scoped impact of interventions may explain why participants felt more motivated and comfortable making manual edits in \chain (Figure~\ref{fig:log_distribute}A).
Nine people felt more compelled to enact controls on sub-tasks, knowing that they did not have to worry about unintended effects on other parts.

Four of them further noted that this clean separation would be tedious (if not impossible) in \sandbox, hence the differences in the perceived controllability in Figure~\ref{fig:likert}.
For example, P13 in \taskReview attempted to replicate the exact same Chain in \sandbox.
They manually divided the original paragraph into three problems, then asked the model for suggestions for each, and to compose the final paragraph.
However, rather than storing suggestions externally and starting fresh for each problem, they simply stacked them together in a single prompt: \exinline{Original paragraph:...; Problem: too much text; Suggestions: 1)...; Problem: Split...}
The resulting long and intertwined text became overwhelming:
\quoteinline{I was very nervous to edit anything, because I didn't know how that was going to impact the end task goals.}

Beyond staged interventions, staged outputs also provided participants with the opportunity to evaluate and improve individual components irrespective of previous failure~\cite{nushi2017human}.
Three participants praised the ability to ``freeze'' their preferred intermediate data points: \quoteinline{I reached some point of some progress in the middle of the Chain and if this works, then it’s fixed when I play with the next step. It doesn't get lost --- unlike the sandbox, where whenever I change something somewhere the result will be completely different} (P10).
Their observations are also in line with the crash-and-rerun capability of crowdsourcing~\cite{little2010turkit}, where local reruns are desirable without affecting previous stages.

%%%%%%%%%%%%%%%%%%%%%%%%%%%%%%%%%%%
\subsubsection{Surfacing the Chaining structure increased transparency.}
\label{sssec:transparent}
\mbox{}\\
%The per-operation prompt structures and their Chaining seem to enrich ``communications'' between humans and the model.
Chaining enriched system transparency, which helped participants better calibrate their expectations of the model.
% when they \emph{review} LLM outputs per step.
As each step of the Chain had a specific role (\emph{Ideation}, \emph{Rewriting}, etc.), they helped narrow the scope of the model's intended functionality, making it easier for participants to understand what to expect from a model that might otherwise seem all-encompassing. 
Nine participants noted this benefit of calibrated expectations. 
For example, P6 commented that \quoteinline{Chaining helped you speak the language. It lift[ed] up the hood and showed you the steps and what's happening at different phrases,} and P15 stated that \quoteinline{having default settings like your templates gave me an idea of how it works.}
As elaborated in Section~\ref{sssec:control}, having isolated steps, each with a reduced scope, also enabled users to better anticipate the potential impact of their inputs, further increasing system transparency.

More globally, Chaining enabled users to develop a more accurate mental model of the LLM's capabilities, by allowing them to tinker with sub-components in a modular and comparative manner.
%Moreover, the modular iterations create implicit contrasting conditions that highlight what the model can or cannot do.
Users could, for example, compare parallel paths to deduce how the model would respond to alternative inputs.
In the \taskFlashcard task, P8 noticed during the \emph{Ideation} step that the model generated more useful English sentences when the \colordot{cinteraction}{types of interactions} was \exinline{accommodation,} compared to \exinline{topics related to public transportation.}
This hinted at the model's better performance when presented with a useful keyword. 
Modifying the order of LLM steps also enabled users to learn aspects of the model's strengths and weaknesses.
When customizing the Chaining structure, five participants tried adding another \emph{Rewriting} step either after the final paragraph (at the end of the Chain), or on the individual presentation problems (early in the Chain).
Though initially unaware that LLMs can suffer from exposure bias (see
C.2), participants quickly discovered through this comparison that the model could more effectively modify \emph{sentences} than \emph{paragraphs}.
This comparison was rare in \sandbox, as it was not obvious to participants that they could keep the LLM functionality but shorten the input.
%As P13 put it, \quoteinline{had I not done the tutorial before this, I would have been able to figure out Chaining, but I wouldn't have been able to figure out Sandbox.}

%Second, across steps, the intermediate data help manage the task completion progress, further increasing system transparency.
%Compared to \quoteinline{blindly aggregating information} (P20), participants especially liked that they could \emph{choose} what data points are relevant to a particular, isolated step, and thereby \quoteinline{enforce that only the intermediate steps that I liked were included later.} (P18).
%Three participants described the different between \sandbox and \chain as bare CLI (command line interface) and IDE (Integrated development environment).
%Quoting P20, \quoteinline{The Chaining experience very much felt like an IDE that enables me to zoom in on individual details, solve sub-problems and then zoom back out to proceed on processing a larger problem}.

%%%%%%%%%%%%%%%%%%%%%%%%%%%%%%%%%%%
\subsubsection{Surfacing the Chaining structure increased debuggability.}
\label{sssec:debug}
\mbox{}\\
The increased transparency in \chain also gave users better debugging mechanisms. %when the model output failed to match expectations. 
When the model output was inconsistent with user intent, participants were at a loss for what to try next in \sandbox.
Because users could conceivably type and modify any natural language prompt in the text box, the scope for ``debugging'' was too expansive. 
% Because users could conceivably type and modify any natural language prompt in the text box, the space of possibilities for ``debugging'' was quite large. 
P9 remarked that \quoteinline{too much freedom can be a curse,} while P7 felt like \quoteinline{sitting down in front of the controls of an airplane, all the knobs are there but I don't know what to do with them.}
Instead, \chain exposed intermediate knobs that helped participants draw a more direct connection between observed model deficiencies, and possible remediation.
P9 found it easier to debug by modifying the inputs and outputs for each step of the Chain, rather than merely re-running the model in \sandbox repeatedly, in the hopes of more promising model results (\quoteinline{I had to constantly delete and rerun things.}).
This may explain why the frequency of \texttt{UNDO} actions was reduced in \chain (Figure~\ref{fig:log_distribute}B).

Accordingly, three interesting debugging mechanisms emerged:
First, the isolated steps in Chaining acted as AI ``unit tests" that enabled users to pinpoint a seemingly global error to its local cause.
For example, participants in \taskFlashcard frequently removed topics irrelevant to traveling (\eg education), so that sub-optimal solutions would not be fed into subsequent steps. 
Second, the ability to create parallel paths and alternate step orders (elaborated in Section~\ref{sssec:transparent}) enabled \emph{comparative debugging}.
%, \ie to highlight breakpoints by comparing alternatives.
Revisiting the case mentioned above,
observing a higher-quality path (\eg using a simple keyword in the prompt like \exinline{accommodation}) helped participants infer how to improve prompts in other parts of the Chain (\eg changing \exinline{topics related to public transportation} to \exinline{public transportation.})

Finally, the ability to propagate a change throughout the entire Chain gave users \emph{immediate feedback on whether a fix was successful}, thereby shortening feedback and iteration cycles.
For example, P3 renamed \colordot{cinteraction}{types of interactions} with \colordot{cinteraction}{places where conversation might occur}, so as to \quoteinline{have flashcards grouped by happening at the airport, restaurant, while walking around streets.}
They were impressed by the changes propagating to the final results: \quoteinline{you can just change a step without affecting other steps but then your final results are reshaped based on that. I didn't think that was going to work that simply.}
This combined ability to both isolate and propagate interventions was key to increasing AI debuggability.
%For example, in \taskFlashcard, P1 changed the prefix \colordot{cen}{Example in English} to \colordot{cen}{Questions in English} once they realized after one running step that they preferred all the flashcard examples to be questions. 

%%%%%%%%%%%%%%%%
\subsubsection{{Scoped objectives in sub-tasks served as guardrails against LLM-inspired tangents}}
\label{sssec:think}
\mbox{}\\
One challenge that hindered participants' performance on the tasks was LLMs' randomness and creative surprises.
The model would often produce outputs that were compelling in their own right, which in turn would derail people from the intended task.
For example, P5 in \taskFlashcard was intrigued by an LLM-generated English sentence, \exinline{That man is suspicious to me,} and started tricking the model into writing a story --- \quoteinline{I want to know what happened to the suspicious man!}
Five out of twenty people wandered from their task goal in \sandbox and began exploring tangents or attempting to ``break'' the model. They had to be reminded several times to get back on track.
Participants later recalled their habit of drifting: \quoteinline{I tried a lot of cool things, but it's not the task I want to complete} (P17).

Interestingly, we found \chain acted as a safeguard against model-inspired tangents, not only because each step of the Chain defined a clear goal, but also because the interconnected data layers motivated participants to deliberately steer outputs of each step away from cascading errors (\eg incorrect problem extraction in the first step of Figure~\ref{fig:flow_review}$b_1$ could lead to a poor final paragraph).
In the \emph{Ideation} steps, participants would even manually move model output around to make sure they fit the topic (P7: \quoteinline{this isn't really about asking for directions, I should put it in accommodation.})
%Participants further leveraged the cascading effect to quickly experiment with alternative prompts.
%For example, in \taskFlashcard, P3 changed the \colordot{ccity}{city to visit} from ``Paris'' to ``Beijing'' and \colordot{cfr}{Examples in French} to \colordot{cfr}{Examples in Mandarin} to easily customize their destination-of-interest;
%Meanwhile, P1 changed the prefix \colordot{cen}{Examples in English} to \colordot{cen}{Questions in English}, which drove the downstream steps to produce all the flashcards in the format of questions.
Ultimately, participants treated the entire task more carefully (see Figure~\ref{fig:likert}, \emph{think through}) --- \quoteinline{if I was trying to do it with speed, I might find the sandbox easier; but if I want to do it with precision, I prefer the Chaining structure.} (P13).

%%%%%%%%%%%%%%%%%%%%%%%%%%%%%%%%%%%
\subsubsection{Additional challenges}
\label{sssec:challenge}
\mbox{}\\
Chaining brought many benefits to human-AI collaboration, but it also presented several challenges.
Nine participants noted that although they found the Chains to be transparent, rich, and educational, they were also more complex, with steeper learning curves. Moreover, while Chaining enabled participants to zoom into sub-tasks in modular ways, it also occasionally made the larger picture more difficult to recall:
Four participants had questions about \quoteinline{how my particular change to this data entry will affect the final result} in \chain (P2), and commented that the end-to-end aspect of \sandbox enabled them to see the direct effects of their actions.
These challenges may have been a side-effect of participants using pre-defined Chains, which may not necessarily reflect their own intuition of how they would have decomposed the task~\cite{xie2017influential, carroll1988mental}.
Most people had a much more fluent experience with the Chains they modified --- \quoteinline{I liked creating my framework.} (P13).
Though beyond the scope of this paper, this raises the question of how to support users in not just using Chains, but also authoring their own Chains, to improve user agency and intuitiveness of Chaining~\cite{wu2022promptchainer}.
%support appropriate task decomposition, so to most effectively utilize the idea of Chaining without causing additional mental burden or task distortion.
%In particular, we should best support users in creating these Chains, to improve their sense of agency.

%Quoting P6, \quoteinline{I feel more at liberty to play with language outside the construct of the Chain.}

Moreover, while Chaining provided better guardrails for staying on task, it may come at the expense of a decreased ability to explore freely; three participants mentioned they would prefer \sandbox for \quoteinline{trying out random things and see if the model can cope} (P3), and \quoteinline{I feel more at liberty to play with language outside the the Chain} (P6). They suggested they would prefer a combination of both systems: \quoteinline{when there's more ambiguity I prefer the sandbox to explore first, but once I have a clear goal, I would use the Chaining to steer myself towards a fixed number of function blocks.} (P13)
 
Inspired by these concerns, we envision future research to focus on relaxing certain structural constraints and providing guidance on LLM Chain creation and refinement, which we detail later in Discussion (Section~\ref{sec:discussion}).

%%%% eval_user.tex ends here %%%%

%%%% eval_case.tex starts here %%%%

\section{Case Studies}
\label{sec:case-study}

%Our user studies show how pipelining support more effective collaboration.

Beyond the user study tasks, LLM Chaining has the potential to enable a wide range of complex applications. 
We illustrate how Chaining could support more diverse applications through two case studies in the domains of software development and accessibility, \added{using the same model in our user study.}
%Second, in the case of assisted text entry, we demonstrate how LLMs could enable if-this-then-that type branching logic in dynamic end-user applications.

\begin{figure*}[t]
\centering
\includegraphics[trim={0 14.2cm 36cm 0cm}, clip, width=0.93\linewidth]{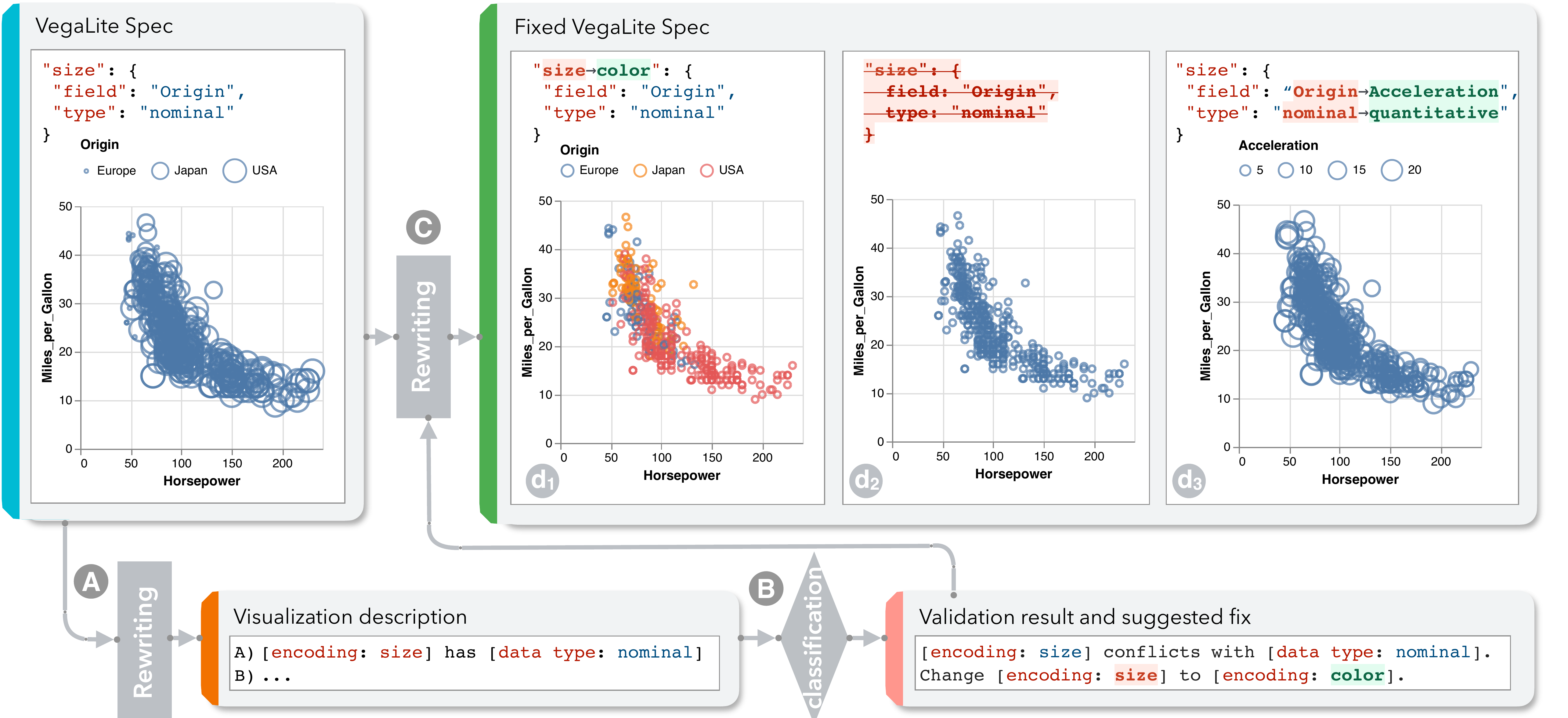}
\vspace{-5pt}
\Description{An example for Chaining-based VegaLite bug fixing.}
\caption{
An example for Chaining-based VegaLite bug fixing
(simplified; the full Chain is in Appendix~\ref{sec:appendix-case-study}).
(A) We first \textbf{rewrite} the \colordot{cvegalite0}{JSON format specs} into \colordot{cdescribe}{natural language descriptions} to make it more parsable, then
(B) \highlight{\textbf{classify} the descriptions to \colordot{cvalidresult}{validate design constraints and suggest fixes}}, and 
(C) finally \textbf{rewrite} the \colordot{cvegalite1}{final spec} based on the suggested fix.
While the LLM generates the fix in $d_1$, users may also choose $d_2$ and $d_3$, both of which can fix the \colordot{cvalidresult}{validated issue} just as effectively.
%Second, there is usually no single gold fix for a given visualization deficiency 
%As such, the intermediate data point, \colordot{cvalidresult}{validation description} on the error (B), becomes even more essential than the final output, \colordot{cvegalite1}{the fixed VegaLite spec} (C).
%While $c_1$ follows the LLM-proposed switch of encoding from \emph{size} to \emph{color} in (B), given the problem of ``[encoding: size] conflicts with [data type: nominal]'', a visualization creator can easily create $c_2$ and $c_3$ by deleting the size encoding entirely, or switching the encoded data field, respectively.
}
\vspace{\anon{-12pt}{-5pt}}
\label{fig:flow_vegalite_simple}
\end{figure*}

\subsection{Case 1: Visualization code debugging}
\label{ssec:viz-debug}
In this case study on visualization code debugging, we uncover how intermediate data points in a Chain can become useful, especially when the end goal of the task is unclear. %We show how Chaining could be used to detect and fix errors in VegaLite specifications, the challenge is the lack of a single right answer.
%A unique challenge here is the lack of a single correct solution: 
Unlike typical code syntax errors, when a visualization violates design constraints~\cite{moritz2018formalizing}, there are usually multiple valid solutions that cannot be objectively ranked.
For example, the \colordot{cvegalite0}{original visualization} (using VegaLite specifications~\cite{satyanarayan2016vega}) in Figure~\ref{fig:flow_vegalite_simple} has a single violation, \ie circle size is continuous and thus should not be used to represent the discrete (nominal) field ``Origin.'' 
However, there may be multiple ways to resolve the issue~\cite{chen2021vizlinter}, such as using color instead of size ($d_1$), removing size information altogether ($d_2$), or changing the data encoded to a continuous ``Acceleration'' field ($d_3$).
Thus, LLMs should reason about the violated constraints for users to adjust the fixes.
%The unclear final fixes mak it important for LLMs to reason about the violated constraints.
%Ideally, LLMs would properly reason about the violated constraint, even when it cannot produce the most desirable fix.
However, in a single run of an LLM, this reasoning can be
challenging, as LLMs have trouble parsing visualization specs in JSON
formats (see LLM  Challenge C.3 in Section~\ref{ssec:llm-challenge}).

We thus created a Chain (see Figure~\ref{fig:flow_vegalite_simple}) that 
(A) \emph{rewrites} the JSON format in natural language,
(B) \emph{classifies} and validates the descriptions, and
(C) \emph{rewrites} the spec.
To explore how the Chain performs in practice, we took examples from VizLinter~\cite{chen2021vizlinter}, used five pairs of erroneous and fixed specs as few-shot prompt examples, and tested the Chain on another five cases. 
One author with sufficient visualization knowledge determined that the Chain correctly revealed the violated constraints for all the test cases, and provided useful fixes for two of them.
We also tried running a single pass of the LLM for comparison on the same examples, using multiple prompt designs.
We observed that output from the single-passes tended to be consistently worse, with at most one correct reasoning. 
This is possibly due to parsing difficulty (see LLM Challenge
C.3), as well as the inability to disentangle the
sub-tasks of validation and rewriting (C.1).
%When directly going from the \colordot{cvegalite0}{original spec} to the \colordot{cvegalite1}{fix}, the LLM would generate unchanged JSON spec, 
%again possibly confused by the parsing difficulty (\ref{challenge:prompt}) and the intertwined tasks of validation \emph{and} rewrite (\ref{challenge:reasoning}).
%When we prompted the LLM to sequentially generate multiple pieces of information in one run (the \colordot{cdescribe}{description}, \colordot{cvalidresult}{validation}, and then \colordot{cvegalite1}{the final fix}), the exposure bias (\ref{challenge:exposure-bias}) would cause it to deviate. 
In contrast, each Chain step was highly scoped, increasing the chance that the intermediate data would be correct.

%Noticeably, when directly transforming from the erroneous specification and the constraints to the correct spec, the LLM either did not recognize the incorrect line at all, or did not make any changes.
%We received the best result when including all of the data layers in Figure~\ref{fig:flow_vegalite_simple} in one multi-data prompt, but it was still limited to one correct reasoning and fix.
%This demonstrates that removing unnecessary intermediate data from later steps is necessary.

%%%%%%%%%%%%%%%%%%%%%%%%%%%%%%%%%%%%%%%

\subsection{Case 2: Assisted Text Entry}
\label{ssec:shothand-expand}

We further demonstrate how Chaining could enable the branching logic in \emph{assisted text entry}.
This is based on a real industry use case that aims to speed up gaze input by requiring fewer character inputs~\cite{majaranta2007text, rough2014evaluation, adhikary2021accelerating}. 
Ideally, a user (\eg person using Alternative and Augmentative Communication technology) would express a full sentence through short abbreviations that an LLM would automatically expand. 
However, there are too many possible expansions to disambiguate, \eg  \exinline{LTSGCHKITOT} could mean \exinline{Let's go check it out
,} \exinline{Let's get coffee and have a chat
,} \exinline{Let's get some chicken in the old town,} etc.
%In fact, even the short \exinline{LTSG} blends \exinline{Let's go,} \exinline{let's get,} etc.
Thus, the end user often needs to resolve the ambiguity or adjust the input.

\begin{figure*}[t]
\centering
\includegraphics[trim={0 30cm 42cm 0cm}, clip, width=0.85\linewidth]{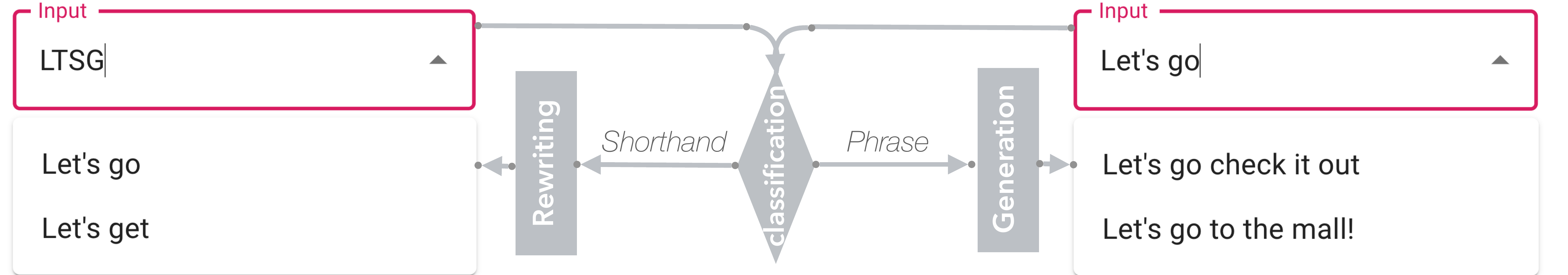}
\vspace{-5pt}
\Description{An example of Chaining-based assisted text entry.}
\caption{ 
An example of Chaining-based assisted text entry (the full Chain is in Appendix~\ref{sec:appendix-case-study}).
To produce better full sentences, we \textbf{classify} the input text to switch between expanding shorthands (through \textbf{Rewrite}) and auto-completing phrases (through \textbf{Generation}).
By wrapping the complex Chaining logic in a simple text field, we provide intuitive interactions for end users.}
\vspace{\anon{-12pt}{-5pt}}
\label{fig:flow_acronym_simple}
\end{figure*}

With Chaining, we enable interactive disambiguation through gradual expansion and if-else logic. 
As shown in Figure~\ref{fig:flow_acronym_simple}, if the user input is a shorthand (\eg \exinline{LTSG}), the LLM should expand it to possible matching phrases (\exinline{Let's go}, \exinline{Let's get}), which the user can select from. 
However, if the input is already a phrase, the LLM should instead auto-complete it (\exinline{Let's go} may trigger \exinline{check it out.})
If the desired option does not appear, the user can also insert additional short-hands for the model to expand again, \eg  \exinline{Let's go CHKITOT}, which would exclude expansions starting with \exinline{Let's get.}
The switch between shorthand expansion and auto-completion
enables better prediction on the full text, which would be nontrivial for a single prompt, given the different natures of the two branches.
This case also provides a glimpse into how LLM Chains can help prototype applications with complex logic but simple interactions (elaborated in the next section).

\section{Discussion \& Future Directions}
\label{sec:discussion}

Our work is a first step towards improving human-LLM interaction through Chaining. 
We found that it not only raises the ceiling of what LLMs can meaningfully support, but also boosts transparency, controllability and debuggability --- key concerns when interacting with generative AI~\cite{amershi2019guidelines, bommasani2021opportunities}. 
\highlight{Interestingly, we achieved this purely by reshaping the interaction mechanism, without any need to retrain the model.
This suggests that LLMs to date may already have the potential to support human-AI collaborations on many complex tasks, if their latent potential can be better realized through thoughtful interaction design.}
Below, we discuss the implications of our studies, as well as future research directions.

%a new paradigm of control that goes beyond a single semantic knob in a single model
\paragraph{Chaining as a new paradigm of control on multiple model units.}
%Just as demonstrated in our use study, LLMs in the sandbox form are capable but passive collaborators, whose (albeit rich and latent) capabilities can only be leveraged through explicit queries.
%They are also easily distracted collaborators, as their generations may be more unpredictable and go off-track, compared to single-purpose AI models~\cite{bommasani2021opportunities}.

%Most of these studies, however, rely on task-specific models, and therefore limit the AI to being a monotonous collaborator that can only make one kind of contribution -- and perhaps not the most effective one.

%While the machine learning or NLP community work towards more advanced and sophisticated modeling, us as the HCI community should further explore how to best expose ``knobs'' in LLMs that end users cab turn, and how to offer hints on when to turn the knobs.

%This opens up new paradigm of control.
%Chaining enables interaction with \emph{multiple} model units.
%These units do not just work together with humans towards distinct sub-goals, but also ``communicate'' through shared inputs and outputs among themselves.
Contrary to recent work in human-AI interaction, which primarily examined how to increase AI controllability through exposing knobs within a model~\cite{louie2020novice, narayan2019similar}, our work opens up the possibility of steering AI using \textit{the model itself} as units to control.
%Contrary to previous work that has primarily examined humans interacting with a single, monotonous model collaborator~\cite{louie2020novice, narayan2019similar}, this setup opens up new opportunities for steering AI:
In other words, beyond controlling properties within a single model unit, users may be able to achieve new kinds of control through manipulating how \textit{multiple} model runs interact with one another, including: how modifications to upstream model units cascade, how to isolate changes between model units, and how to improve user inputs by comparing the effectiveness of parallel model runs. 
As language models grow in size and capability, they may ironically allow users to treat them as smaller entities of abstraction --- serving as building blocks towards larger human goals. 

%As the machine learning community continues to work towards more advanced and sophisticated models, the HCI community could likewise be empowered to innovate new ways to expose such controls to end users, by manipulating what building blocks the model units should serve, and how they can be combined.
%Hence, as language models increase in size and capability, they may ironically allow users to treat them as smaller entities of abstraction -- serving as building blocks towards larger human goals. 

%For example, alternative designs may interleave LLM steps with human-computation steps, taking into consideration which types of LLM operations are most beneficial if done by humans, and which could be better handled by an LLM. 

We envision the HCI community  innovating more types of building blocks that a model can provide, as well as the ways they can be combined. 
In particular, model units could be used not only to accomplish sub-tasks, but also to more thoroughly aid in the task decomposition design and debugging process.
%In particular, model units could be used not only for accomplishing \emph{sub-tasks}, but also for more explicitly aiding in the task decomposition design and debugging process itself. 
To overcome users' own systematic omissions~\cite{wu-etal-2021-polyjuice}, an upstream unit could be designed to help users create sub-tasks to begin with, similar to metaprompting~\cite{reynolds2021prompt}. 
Or, model units could serve as checkpoints along the Chain to ensure data correctness (similar to assertions in code).
Moreover, while the Chains in this paper consisted of only LLM steps, alternative designs may also interleave LLM steps with human-computation steps, depending on which roles each collaborator could best fill.

%we can create \emph{utility LLM units} in addition to \emph{sub-task units}, to assist end users with prompt engineering per sub-task.
%To overcome users' own systematic omissions~\cite{wu-etal-2021-polyjuice}, we may leverage the model-inspired diversity (observed in Section~\ref{sssec:collaboration}), and define a \emph{example sourcing} unit, \ie the LLM itself generates diverse demonstrative examples for instructing another LLM step.
%Similarly, a series of \emph{paraphrasing} steps (\eg use more jargons,  more succinct descriptions) would aid in exploring the alternative prompts available.

%Alternative designs may also interleave LLM steps with human-computation steps, depending on which roles the collaborators fulfill.
%sub-tasks are best handled by humans, and which should be left to an LLM.
%When using AI for creative inspiration, perhaps AI should take the lead, and humans should follow and build on the AI's suggestions~\cite{gero2019metaphoria};
%whereas for decision-making tasks, or those where users wish to retain a strong sense of authorship (\eg co-creating music~\cite{louie2020novice}), postponing AI involvement until after humans have made their own decisions may be more beneficial, as it avoids anchoring or biasing effects~\cite{englich2006playing, bansal2021does, wang2019designing,sheng2021societal}.

\paragraph{Chaining for rapid prototyping of integrated applications}

%y enabling the development of sophisticated models without having to collect significant amounts of data and train large models from scratch. 
%This could enable even non-ML experts to quickly prototype AI-infused applications. At the same time, the powerful generative and potentially multi-modal capabilities of foundation models could offer a far higher “ceiling” [Myers et al. 2000] of what types of interactions are achievable both in terms of their quality and diversity as we will discuss below. However, how successfully we can leverage these capacities will depend on how effectively we can wrangle foundation models into forms that will be more manageable by application developers

% Something specific that i see as neat is that the iteration speed can be super quick for a new kind of interaction. Feels like end user testing + adding features and refining taxonomies can be quicker than ever by many orders of magnitude.
%There's something super compelling and interesting about this approach, even if expensive in TPUs: to be able to define a space of UX and "fake" models for it quickly, it creates a real and compelling experience. I think it may change the way we do prototyping. And then I think the next stage of research is how to turn a set of prompts into something higher quality, presumably creating a larger dataset, and fine tuning smaller models that can run in production.

Chaining also opens up new possibilities for designing AI-infused applications.
%Rapid prototyping is already possible with individual 
With LLMs' easy adaptation to natural language prompts, users could conceivably already prototype custom ML functionality with lower effort, as they bypass the otherwise necessary but expensive process of collecting data and designing models upfront~\cite{bommasani2021opportunities}.
% or design models upfront, as is typically needed to develop machine learning models.
%LLMs already make it conceivable for users to prototype custom ML functionality through natural language prompting, without the need for expensive upfront data collection or model design typical in machine learning development. 
Chaining further accelerates this design process. 
Taking advantage of interactions between multiple LLM steps, developers could build multiple Chains to envision possible flows of how an application may be used, and then perform A/B testing on those Chains.
For example, in the case of assisted text entry (Section~\ref{ssec:shothand-expand}), developers could quickly prototype what might happen if end users were allowed to provide more context: \eg if the user is \exinline{having a meeting in 5 minutes,} then \exinline{Let's go} is more likely than \exinline{Let's get} for the abbreviation \exinline{LTSG.} 
They could test this interaction by adding an additional layer of input to the shorthand expansion step. %of the Chain. 

One might argue that, because each run of an LLM involves some computational overhead, chaining may introduce additional costs that need to be weighed against their benefits. However, as indicated above, a key benefit of chaining is that it could flexibly power a wide range of prototypes and applications, \emph{without} the need to train or build bespoke, single-purpose AIs. Thus, we believe the saved efforts outweigh the cost.
%Or, a designer could add another LLM step of rewriting after the auto-completion step, to test if end users would benefit from paraphrasing or grammar corrections. 
%While each run of an LLM may involve some computational overhead, we believe this gain in comparative prototyping would outweigh the additional costs.
%may introduce additional costs that need to be weighed against their benefits. 

%A key benefit of chaining is that it can flexibly power a wide range of prototypes and applications that involve multiple inherent sub-tasks, \emph{without} the need to build bespoke, single-purpose AIs for each sub-task. 
%The flexibility of natural language, coupled with the modularity of chaining, could enable rapid prototyping of new ML applications without requiring the upfront expensive cost of collecting training data or designing a new model. 
%Moreover, since we can either grant people access to the intermediate steps (as in our user study tasks) or hide them (as in Section~\ref{ssec:shothand-expand}), we can easily switch between development mode (where debuggability is the key) and deployment mode (where intuitive interactions are more important). 
%Of course, each LLM call involves some computational overhead, but we believe the ease of prototyping drastically outweighs the extra costs.

\paragraph{Balancing between structured scaffolding and free exploration}
While Chaining provided guardrails and scaffolding for helping users accomplish the task at hand, it also limited their ability to explore freely. 
Yet, experimenting, tinkering, and interacting are key to users forming mental models for AI~\cite{narayan2019similar}.
One way to balance between structure and exploration is to loosen structural constraints \emph{within} steps. 
For example, it may be useful to permit users to customize prompts within each step in a Sandbox-like environment, and to define their own input and output parsers.
In other words, rather than providing a full implementation of steps, a Chain could define the API with input-output types, and ask users to fill in the implementations for each step. Or, a small Sandbox could be provided along-side the Chaining interface, for users to occasionally use when they need to experiment with a new approach.

Meanwhile, though our studies mostly explored how humans \emph{use} pre-defined LLM Chains, a natural follow-up question becomes whether end users can effectively \emph{author} their own LLM Chains. 
Indeed, one potential downside of Chaining is that it may \emph{decrease} transparency if the pre-built Chain does not match the way a user would naturally break down the task (mentioned in Section~\ref{sssec:challenge}).
We believe our operations can serve as a starting point for future work on authoring.
With the templates, users could instantiate an LLM step by defining the data layers and selecting the operations. 
In our study, most participants were able to spot deficiencies and refine the default Chains accordingly. 
Thus, we envision that a set of generic default Chains could help onboard end users to the idea of LLM Chaining, and inspire them to author more tailored Chains.
We leave end user authoring of Chains to future work.

\paragraph{Enhancing LLM Chain design and refinement}
Our work centered mostly on moderately complex tasks that can be naturally broken down.
However, decomposition might be less straightforward in some cases~\cite{kim2017mechanical}.
Tasks with more complex interdependence may lose coherence and quality if they are split into independent sub-parts.
%For tasks with more complex interdependence, splitting them into independent sub-parts may decrease the coherence and quality of the final results.
For example, in the Review task (Figure~\ref{fig:flow_review}), we treated the different problems independently. 
However, if the problems are interrelated, keeping them together would promote more effective suggestions (\eg not engaging \emph{and} speaks too quietly).
Moreover, while users had the option of excluding specific data layers along the way (\eg the original review in Figure~\ref{fig:flow_review} is not fed into the final step), the information loss may also lead to task distortion or compression~\cite{reynolds2021prompt}.
%How can we help users craft Chains in a way that makes the most of the issues presented by Chaining?
%In light of the challenges Chaining presents, what can we do to help users develop Chains that maximize their utility?
In light of these issues, future work could investigate how to assist users in crafting the steps of a Chain to maximize its utility~\cite{kittur2011crowdforge}. For example, users could be provided strategic guidance on iterative Chain improvements, such as using paired comparisons and version control of Chain edits to help users decide whether to keep or further decompose an existing step.

%%%% discussion.tex ends here %%%%

%%%% conclusion.tex starts here %%%%

\section{Conclusion}

%Although much recent work has examined the algorithmic capabilities of LLMs (large language models), less work has examined the ways in which we can design LLMs to better support human activities. In this project, we start with the discovery that while LLMs tend to do well on simple tasks, many socially beneficial human activities (e.g. education, well-being, productivity) involve complex tasks that cannot be easily handled via a single run of an LLM. To address this, 

In this work, we introduce the notion of ``Chaining'' multiple LLM steps together, such that the output of one step is the input to the next. 
We present an interactive system where users can modify these Chains, along with their intermediate results, in a modular way. 
We find that Chaining not only enhanced the quality of the task outcome, but also improved user satisfaction, with an increased sense of control and collaboration, a greater perception of transparency of the LLM system, and more support of the user's thought processes.
Furthermore, we envision with case studies that LLM Chaining may be advantageous for complex AI-infusion applications and in cases where intermediate reasoning is more important than the final output.
We encourage future work to explore how LLMs can serve other kinds of building blocks, how Chains can be used in rapid prototyping, and strategies that can help users build and iterate on Chains.

%%%% conclusion.tex ends here %%%%

%%%% ack.tex starts here %%%%

\begin{acks}
We gratefully thank 
Shanqing Cai,
David Dohan, 
Aaron Donsbach,
Noah Fiedel,
Anna Huang,
Ellen Jiang,
Ajit Narayanan,
Kristen Olson,
Meredith Ringel Morris,
Adam Pearce,
Jascha Sohl-dickstein,
Edwin Toh,
Subhashini Venugopalan,
and
Google PAIR team for their helpful comments. We also appreciate the valuable input from our study participants.
%, and the constructive comments from the anonymous reviewers.
\end{acks}

%%%% ack.tex ends here %%%%

\balance

\bibliographystyle{ACM-Reference-Format}
\bibliography{ref}%% Commented by merge tool

%%
%% If your work has an appendix, this is the place to put it.
\appendix

%\clearpage\newpage
\onecolumn

%%%% primitive_category.tex starts here %%%%

\section{Identifying LLM Primitive Operations}
\label{sec:appendix-primitive-category}

\begin{table*}%2%[th]
\centering
\begin{tabular}{@{}c@{}}
\includegraphics[width=\textwidth]{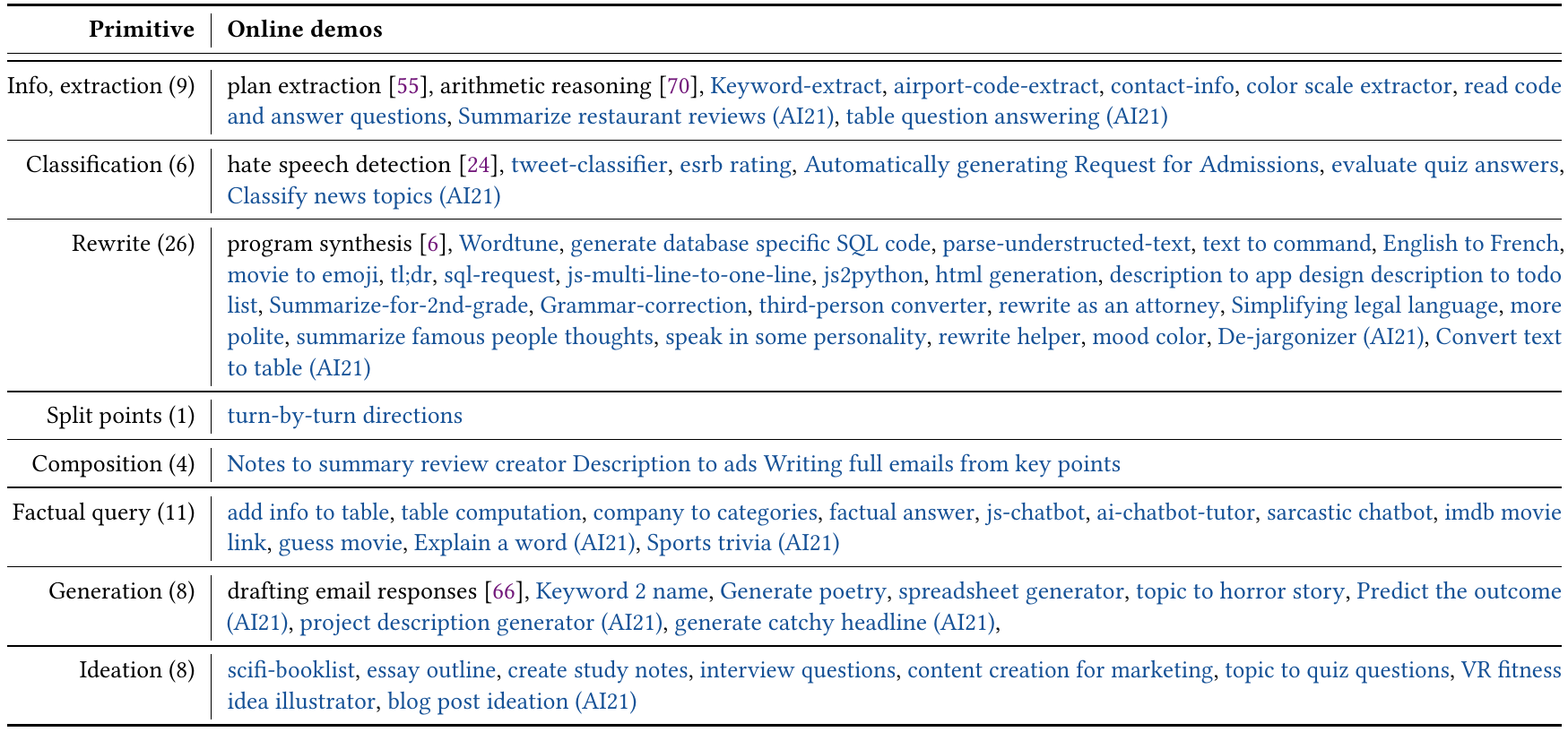}
\end{tabular}
\caption{A survey of 73 online demos that inspired the design of our operation, mostly from published manuscripts, \href{https://beta.openai.com/examples}{the OpenAI official GPT-3 example page}, \href{https://studio.ai21.com/playground}{the AI21 tutorial}, 
and \href{https://github.com/elyase/awesome-gpt3}{the demo collection repository}.
All the links are last accessed in 2021/08.
}
\label{table:primitive_category}
\vspace{-10pt}
\end{table*}

We reviewed 73 existing demos to identify promising LLM capabilities that may help overcome the challenges above by scoping the inputs/outputs to be more amenable to what an LLM can handle. 

First, we collected demos from LLM official websites (\eg GPT-3 and Jurassic), social media, and published case studies by searching for keywords including ``GPT-3,'' ``language model,'' ``prompt,'' etc.
After removing some demos that were highly open-ended rather than targeted (\eg generic chatbots), we iteratively sorted the demos into eight LLM primitive operations, as shown in Table~\ref{table:primitive}. For example, we distinguished between operations that had different expected data mappings (one-to-many v.s. many-to-one), and different application types (deterministic v.s. creative).
We then grouped the primitives into three high level groups based on which LLM challenge they may help address. The groups also appear to be consistent with categories presented on the GPT-3 tutorial page,\footnote{https://beta.openai.com/docs/guides/completion/introduction} which highlighted typical NLP tasks like \emph{Classification}, \emph{Generation} (\ie gather additional information in Table~\ref{table:primitive}b), \emph{Transformation} (\ie re-organization).
Finally, we further refined the primitive categories and names based on feedback from three pilot users (one LLM expert and two UX engineers with basic knowledge of LLM prompting).

%%%% primitive_category.tex ends here %%%%

%%%% app-eval-user.tex starts here %%%%

\section{Additional Details for User Study}
\label{sec:appendix-user-study}

%%%%%%%%%%%%%%%%%%%%%%%%%%%%%%%%%%%%%%%%%%%%%%%%
\subsection{Questions in the Exit Survey}
\label{ssec:appendix-user-study-survey}

After completing the given task in both conditions, participants self-rated their experience on the following dimensions, in the form of seven-point Likert scale~\cite{likert1932technique}.
Each question was asked twice, once on \sandbox and once on \chain.
They described their reasoning along with the ratings. 

%\sherry{Remember to change the name.}

\begin{itemize}
\item \emph{Match goal}: I'm satisfied with my final results from \conditionNameGrp; they met the task goal.
\item \emph{Think through}: The \conditionNameGrp system helped me think through what kinds of outputs I would want to complete the task goal, and how to complete the task.
\item \emph{Transparent}: The \conditionNameGrp system is transparent about how it arrives at its final result; I could roughly track its progress.
\item \emph{Controllable}: I felt I had control creating with the \conditionNameGrp system. I can steer the system towards the task goal. 
\item \emph{Collaborative}: In \conditionNameGrp, I felt I was collaborating with the system to come up with the outputs. 
\end{itemize}

Additionally, participants also answered the following two free form questions:
\begin{itemize}
\item \emph{Difference}: What were the differences, if any, between the experience of completing the task using \sandbox and \chain?
\item \emph{Vision}: If you were using language models in your work, in what situations would you prefer to use \sandbox? \chain? Can you think of 1-3 concrete examples?
\end{itemize}

%%%%%%%%%%%%%%%%%%%%%%%%%%%%%%%%%%%%%%%%%%%%%%%%

\subsection{Clickstream Categorization}
\label{ssec:appendix-user-study-clickstream}
we log the text status before and after each round of model run.
Through sequence match, we recover what's generated by the model after each run, and how the participants edit the text in between of two runs. We split the logs into: (1) \texttt{RUN} the model, (2) \texttt{UNDO} the model, where people removed the generations from the previous run, making the resulting text more similar to prior to the previous run, (3) \texttt{FORMAT}, where people only add or remove line split or formatting-related stopwords, (4) \texttt{CREATE-CONTENT}, where people only insert meaningful spans to the text, (5) \texttt{CURATE-CONTENT}, where people make all the other kinds of refinements on the existing text --- in \chain, this is a merge of changing the instruction, prefix, and the data entries. We also logged (6) \texttt{CHANGE-TEMPERATURE} to denote when people make non-text based change on the model input, \ie temperature.

On top of the logs, we define \emph{consecutive runs} (in Figure~\ref{fig:log_distribute}A) as those in which users did not change anything after the previous run (or only add formatting through line changes or adding stopwords, \ie \texttt{RUN}+\texttt{FORMAT}).
Otherwise, the logs are counted as humans making \emph{edits}.

%%%%%%%%%%%%%%%%%%%%%%%%%%%%%%%%%%%%%%%%%%%%%%%%

\subsection{Case 0: Metaphor Creation (Used in tutorial)}
\label{ssec:appendix-user-study-metaphor}
\begin{figure*}[t]
    \centering
    \includegraphics[trim={0 28cm 44cm 0cm}, clip, width=0.9\columnwidth]{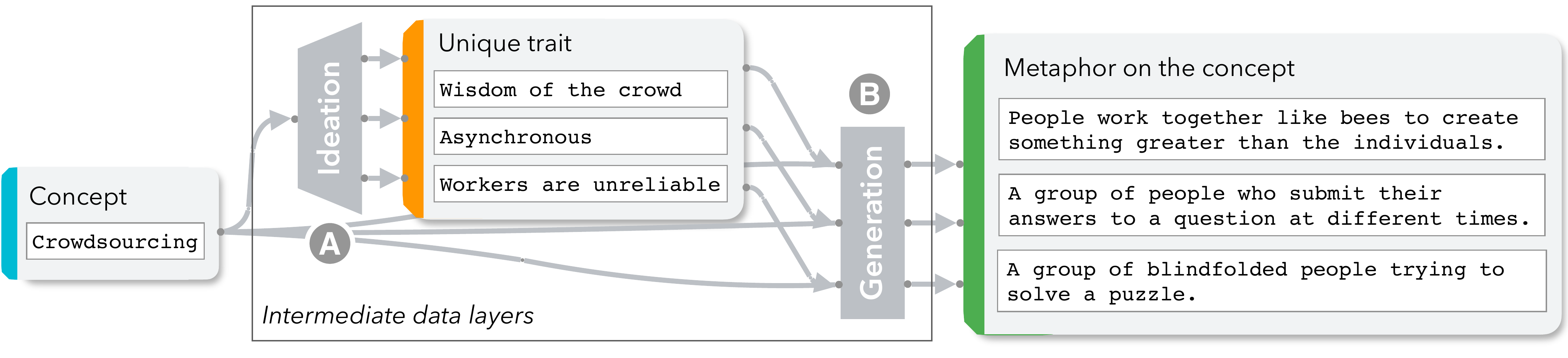}
    \vspace{-5pt}
    \Description{The chain for acronym expansion.}
    \caption{ 
    The pipeline for acronym expansion.
    The steps include:
    (A) An \textbf{ideator} that brainstorms various \colordot{ctrait}{unique traits} for \colordot{cconcept}{the concept} (crowdsourcing). 
    (B) For each trait, a \textbf{generator} creates a related \colordot{cmetaphor}{metaphor}.
    }
    \vspace{-10pt}
    \label{fig:flow_metaphor}
\end{figure*}

\paragraph{\textbf{\emph{Description.}}}
Create metaphors for the concept of \emph{crowdsourcing}, so that we can explain the \emph{different aspects} of crowdsourcing in a poetic way.
The pipeline is as in Figure~\ref{fig:flow_metaphor}.

A metaphor may look like:

\squote{In crowdsourcing, people are like bees; they work together to make honey.}

With the \emph{concept} being ``crowdsourcing'', the \emph{simile} being ``bees'', and the \emph{similar aspect} being ``work together.''

\paragraph{\textbf{\emph{Default baseline commands.}}}
\begin{enumerate}
\item In the form of question answering,
    \taskquote{\tunit{Question: What is a good metaphor for crowdsourcing?\newline
    Answer: }
    {a swarm of bees.}}
\item In the form of command instruction,
    \taskquote{\tunit{Write a metaphor for the concept of crowdsourcing.\newline
    Concept: crowdsourcing\newline
    Metaphor: }{Crowdsourcing is like a game of chess. A crowdsourcer's skills, as in a chess player's skills, are combined with another person's skills to make something new.}}
\item List enumeration
    \taskquote{\tunit{The following is a list of metaphors on crowdsourcing.\newline
    1. Crowdsourcing is like
    }{a beehive -- Many people (bees) contribute to a larger cause.}}
\item Few-shot example,
    \taskquote{\tunit
    {Concept: gratitude\newline
    Metaphor: gratitude is like a stream in that it's a force that can carry you along.\newline
    \#\#\#\newline
    Concept: loss\newline
    Metaphor: loss is like a wing in that it's something you never wanted to lose, and it can take you away.\newline
    \#\#\#\newline
    Concept: crowdsourcing\newline
    Metaphor:}{crowdsourcing is like a team sport in that it brings people to achieve one goal.}}
\end{enumerate}

\subsection{Case 1: Flashcard Creation}
\label{ssec:appendix-user-study-flashcard}

\paragraphBold{Description.}
Suppose you will be traveling to Paris next week, and you would like to create flashcards to learn about some basic French so you can have basic conversations with local people whenever you are in a non-English speaking region.
Your goal is to create flashcards that are both \textbf{diverse} and \textbf{personalized} to your travel desires.

A flashcard may look like:

\squote{English: Where is a good restaurant?; 
French: Où est un bon restaurant?}

\paragraphBold{Default baseline commands.}
we offer three versions:

\begin{enumerate}
\item In the form of question answering,
    \taskquote{\tunit{Question: What are some English and French sentence pairs useful for traveling to Paris?\newline
    Answers:\newline}
    {English: Where is a good restaurant?\newline
    French: Où est un bon restaurant?}}
\item In the form of list enumeration,
    \taskquote{\tunit{The following are English and French sentence pairs useful for traveling to Paris.\newline
    1. }{Que fais-tu ici? / What are you doing here?}}
\item Few-shot example,
    \taskquote{\tunit
    {English: I do not speak French.\newline
    French: Je ne parle pas français.\newline
    \#\#\#\newline
    English: Where is a good restaurant?\newline
    French: Où est un bon restaurant?\newline
    \#\#\#\newline}{English: Where's the bus station?\newline
    French: Où est la gare routière?}}
\end{enumerate}

%%%%%%%%%%%%%%%%%%%%%%%%%%%%%%%%%%%%%%%%%%%%%%%%
\subsection{Case 2: Peer Review Paragraph Rewriting}
\label{ssec:appendix-user-study-review}

\paragraph{\textbf{\emph{Description.}}}
The following is a peer review paragraph on Alex's presentation. 
Please rewrite it to be more \textbf{friendly and constructive}. The paragraph should include \textbf{1-3 concrete and actionable suggestions} for \textbf{each} of Alex’s presentation problems.Original feedback:

\squote{Alex could improve his presentation skills. He has too much text on his slides. His presentation meanders from topic to topic without a clear structure. He also does not engage with his audience when he presents.}

An example output from the language model:

\squote{Alex, you have a lot of good ideas in your presentations. However, your slides are too cluttered with text. I would recommend that you use more graphics, pictures, and white space. Your slides should be simple enough that you could read your presentation off of the slides.  This will allow you to engage the audience and not have to read from your notes. Also, you need to have a clear structure to your presentation. Lastly, you need to practice speaking. This will familiarize you with the material and reduce the likelihood of you meandering.}

\paragraph{\textbf{\emph{Default baseline commands.}}}
we offer two versions:

\begin{enumerate}
\item End-to-end version,
    \taskquote{\tunit{Given the feedback, rewrite it into a friendly paragraph with concrete suggestions on each of Alex’s presentation problems.\newline
    Feedback: Alex could improve his presentation skills. He has too much text on his slides. His presentation meanders from topic to topic without a clear structure. He also does not engage with his audience when he presents.\newline
    Friendly paragraph: }
    {\emph{[\modelname generation]}}}
\item Two-step version, where we query \modelname for improvement suggestions first, and then ask it to integrate the problem and the suggestion.
    \taskquote{\tunit{Alex could improve his presentation skills. He has too much text on his slides. His presentation meanders from topic to topic without a clear structure. He also does not engage with his audience when he presents.\newline
    Give Alex some suggestions on his presentation:\newline
    1. }
    {\emph{[\modelname generation]}}\newline
    \tunit{Write one friendly paragraph that covers all the presentation problems and suggestions:}
    {\emph{[\modelname generation]}}
    }
\end{enumerate}

%%%% app-eval-user.tex ends here %%%%

%%%% app-eval-case.tex starts here %%%%
\clearpage

\section{Full LLM Chains for Case Studies}
\label{sec:appendix-case-study}

\begin{figure*}[h]
\centering
\Description{The LLM Chain for visualization bug fixing (in VegaLite).}
\includegraphics[trim={0 5cm 37cm 0cm}, clip, width=1\linewidth]{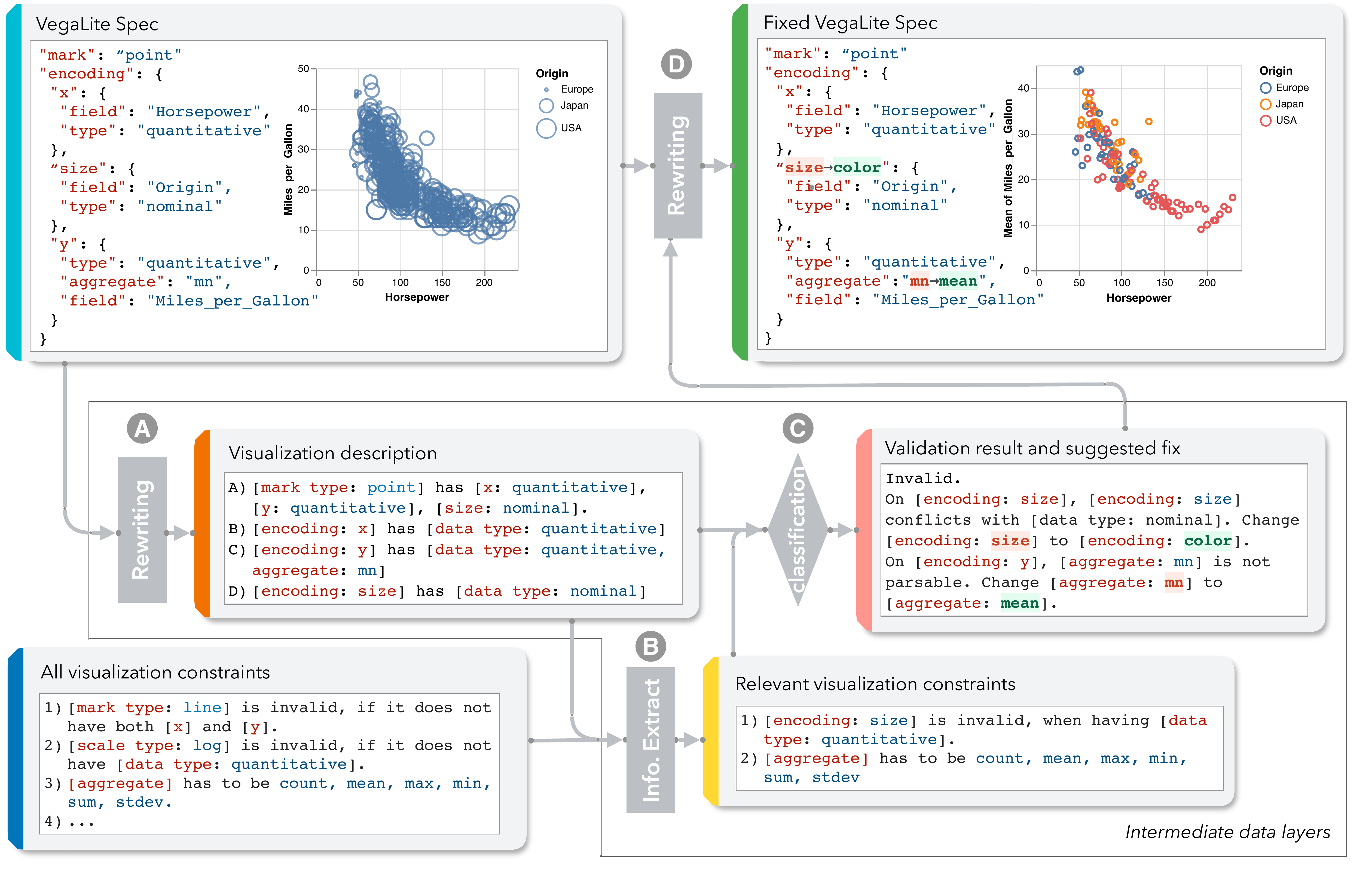}
\vspace{-20pt}
\caption{ 
The LLM Chain for visualization bug fixing (in VegaLite).
The stages include:
(A) A \textbf{Rewrite} step that transforms the \colordot{cvegalite0}{json format VegaLite spec} into \colordot{cdescribe}{natural language description}, so to eliminate noise from data. 
%Prompt (through few-shot): 
%\tprompt{``Extract a list of \textcolor{cdescribe}{information blocks} from the given \textcolor{cvegalite0}{VegaLite code}.''}
(B) An \textbf{Information Extraction} step that locate \colordot{crelaterule}{related visualization rules}.
%Prompt (through few-shot): 
%\tprompt{``Given the \textcolor{cdescribe}{information description block}, extract \textcolor{crelaterule}{related visualization rules} from \textcolor{callrule}{all the possible rules}.''}
(C) A \textbf{Classification} step that verifies the description as either \colordot{cvalidresult}{valid or invalid} (with concrete errors and fixes).
%Prompt (through few-shot): 
%\tprompt{``This is a verifier that checks if the \textcolor{cdescribe}{description} is \textcolor{cvalidresult}{valid} based on the \textcolor{crelaterule}{given rules}, and explain.''}
(D) A \textbf{Rewriting} step that generates \colordot{cvegalite1}{the fixed VegaLite spec} based on the \colordot{cvalidresult}{validity reasons}. 
}
%\vspace{-10pt}
\label{fig:flow_vegalite}
\end{figure*}

\begin{figure*}[h]
\centering
\includegraphics[trim={0 24cm 39cm 0cm}, clip, width=1\linewidth]{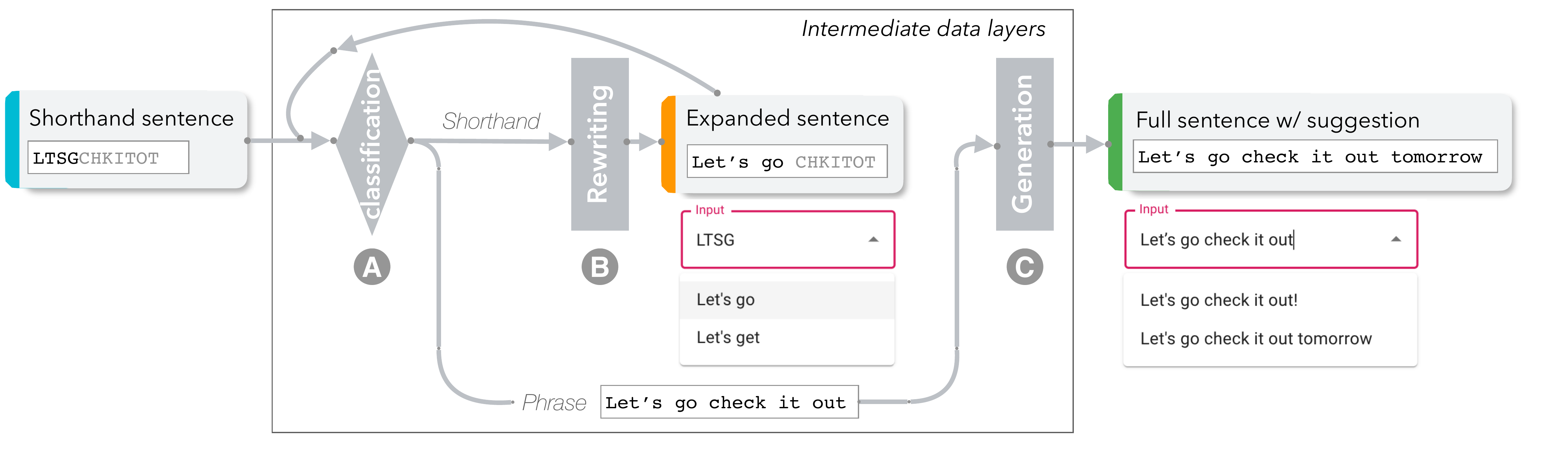}
\Description{The full LLM Chain for assisted text entry.}
\vspace{-20pt}
\caption{ 
The LLM Chain for assisted text entry.
The stages include:
(A) A \textbf{Classification} steps that detects whether a \colordot{cacronym}{given sentence} \colordot{chasacronym}{contains a shorthand or not}. 
(B) If there exists certain shorthand, a \textbf{Rewriting} step expands it, so we arrive at the \colordot{cexpandcent}{expanded sentence} which can become the context for additional shorthand inputs.
For ``LTSG'', it can be \exinline{Let's go} or \exinline{Let's get}, which relies on human selection.
(C) Otherwise, a \textbf{Generation} step autocompletes \colordot{cfullsent}{the sentence}.
}
\vspace{-10pt}
\label{fig:flow_acronym}
\end{figure*}

%%%% app-eval-case.tex ends here %%%%

%%%% primitive_operation.tex starts here %%%%

\clearpage
%\newpage
\section{The Full Implementation of Primitive Operations}
\label{sec:appendix-primitive-implement}

%%%% figures/primitives_full.tex starts here %%%%

\begin{table*}[ht]%3%
\begin{tabular}{@{}c@{}}
\includegraphics[width=\textwidth]{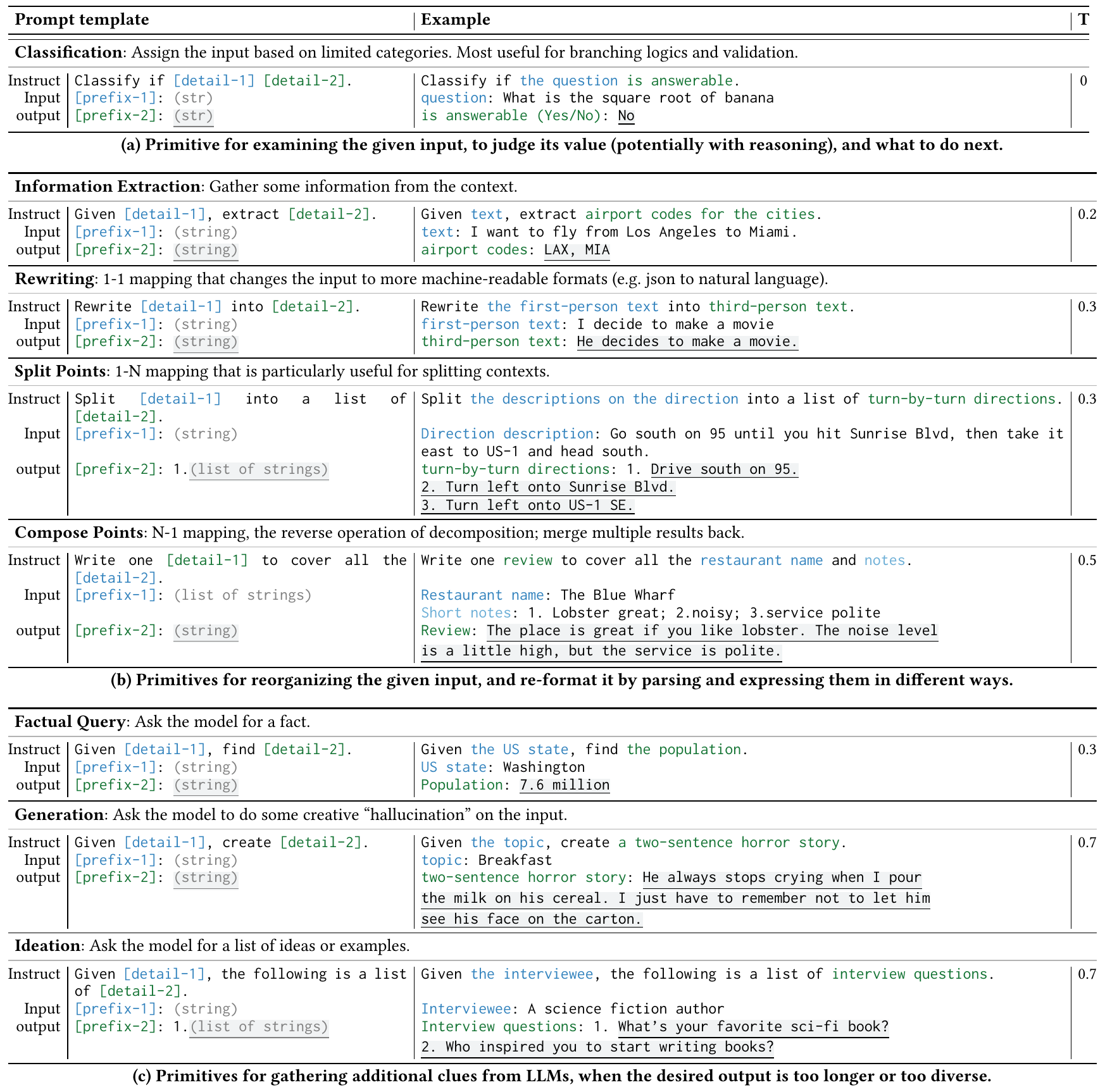}
\end{tabular}
\caption{We design a list of primitive building blocks, each with default prompting templates and temperatures, and group them by their intended objectives. Examples are taken from \url{https://beta.openai.com/examples}.
}
\label{table:primitive_full}
\vspace{-10pt}
\end{table*}

%%%% figures/primitives_full.tex ends here %%%%

%%%% primitive_operation.tex ends here %%%%

\end{document}